\documentclass[12pt,letterpaper]{article}
\usepackage{epsfig,rotating,setspace,latexsym,amsmath,epsf,amssymb,bm,theorem}
\usepackage{cite,appendix}

\title{Secrecy in Cooperative Relay Broadcast Channels\thanks{This
work was supported by NSF Grants CCF 04-47613, CCF 05-14846, CNS
07-16311 and CCF 07-29127, and presented in part at the
Information Theory and Applications Workshop, San Diego, CA,
January 2008 and the IEEE International Symposium on Information
Theory, Toronto, Canada, July 2008~\cite{ISIT_08}.}}

\author{Ersen Ekrem \qquad Sennur Ulukus\\
\normalsize Department of Electrical and Computer Engineering\\
\normalsize University of Maryland, College Park, MD 20742\\
\normalsize {\it ersen@umd.edu} \qquad {\it ulukus@umd.edu} }

\theorembodyfont{\normalfont}

{\theoremstyle{plain} \newtheorem{Remark}{Remark}}
\newtheorem{Theo}{Theorem}
\newtheorem{Prop}{Proposition}

\newtheorem{Cor}{Corollary}

\newenvironment{proof}[1]{\medskip\par\noindent
{\bf Proof:\,}\,#1}{{\mbox{\,$\blacksquare$}\par}}
\begin{document}
\maketitle

\begin{abstract}
We investigate the effects of user cooperation on the secrecy of
broadcast channels by considering a cooperative relay broadcast
channel. We show that user cooperation can increase the achievable
secrecy region. We propose an achievable scheme that combines
Marton's coding scheme for broadcast channels and Cover and El
Gamal's compress-and-forward scheme for relay channels. We derive
outer bounds for the rate-equivocation region using auxiliary
random variables for single-letterization. Finally, we consider a
Gaussian channel and show that both users can have positive
secrecy rates, which is not possible for scalar Gaussian broadcast
channels without cooperation.
\end{abstract}

\newpage

\section{Introduction} \label{sec:Introduction}
The open nature of wireless communications facilitates cooperation
by allowing users to exploit the over-heard information to
increase achievable rates. However, the same open nature of
wireless communications makes it vulnerable to security attacks
such as eavesdropping and jamming. In this paper, we investigate
the interaction of these two phenomena, namely cooperation and
secrecy. In particular, we investigate the effects of cooperation
on secrecy.

The eavesdropping attack was first studied from an information
theoretic point of view by Wyner in \cite{Wyner}, where he
established the secrecy capacity for a {\it single-user} {\it
degraded} wire-tap channel. Later, Csiszar and Korner
\cite{Korner} studied the general, not necessarily degraded, {\it
single-user} eavesdropping channel, and found the secrecy
capacity. More recently {\it multi-user} versions of the secrecy
problem have been considered for various channel models.
References \cite{Aylin_Cooperative, Ruoheng_4,
Yingbin_1,Ruoheng_3,Ersen_Sennur} consider multiple access
channels (MAC), where in \cite{Aylin_Cooperative, Ruoheng_4} the
eavesdropper is an external entity, while in
\cite{Yingbin_1,Ruoheng_3,Ersen_Sennur} the users in the MAC act
as eavesdroppers on each other. References \cite{Ruoheng2,
Ruoheng} consider broadcast channels (BC) where both receivers
want to have secure communication with the transmitter; in here as
well, each receiver of the BC is an eavesdropper for the other
user. References \cite{Aylin_Yener, Oohama,He_ISIT, Hesham,
Melda_1,Bloch_Relay} consider secrecy in relay channels, where in
\cite{Aylin_Yener, Oohama,He_ISIT}, the relay is the eavesdropper,
while in \cite{Hesham, Melda_1} there is an external eavesdropper.
In \cite{Bloch_Relay}, the relay helps the transmitter to improve
its rate while it receives confidential messages that should
be kept hidden from the main receiver.

In a wireless medium, since all users receive a version of all
signals transmitted, they can cooperate to improve their
communication rates. The simplest example of a cooperative system
is the relay channel \cite{Cover} where the relay helps increase
the communication rate of a single-user channel using its
over-heard information. Multi-user versions of cooperative
communication have been studied more recently. In
\cite{sendonaris}, a MAC is considered where both users over-hear
a noisy version of the signal transmitted by the other user, and
transmit in such a way to increase their achievable rates. In
\cite{Kramer, Servetto, Veeravalli}, cooperation is done on the
receiver side, where in a BC, one or both of the receivers
transmit cooperative signals to improve the achievable rates of
both users.

\begin{figure}[t]
\begin{center}
\epsfxsize = 12 cm  \epsffile{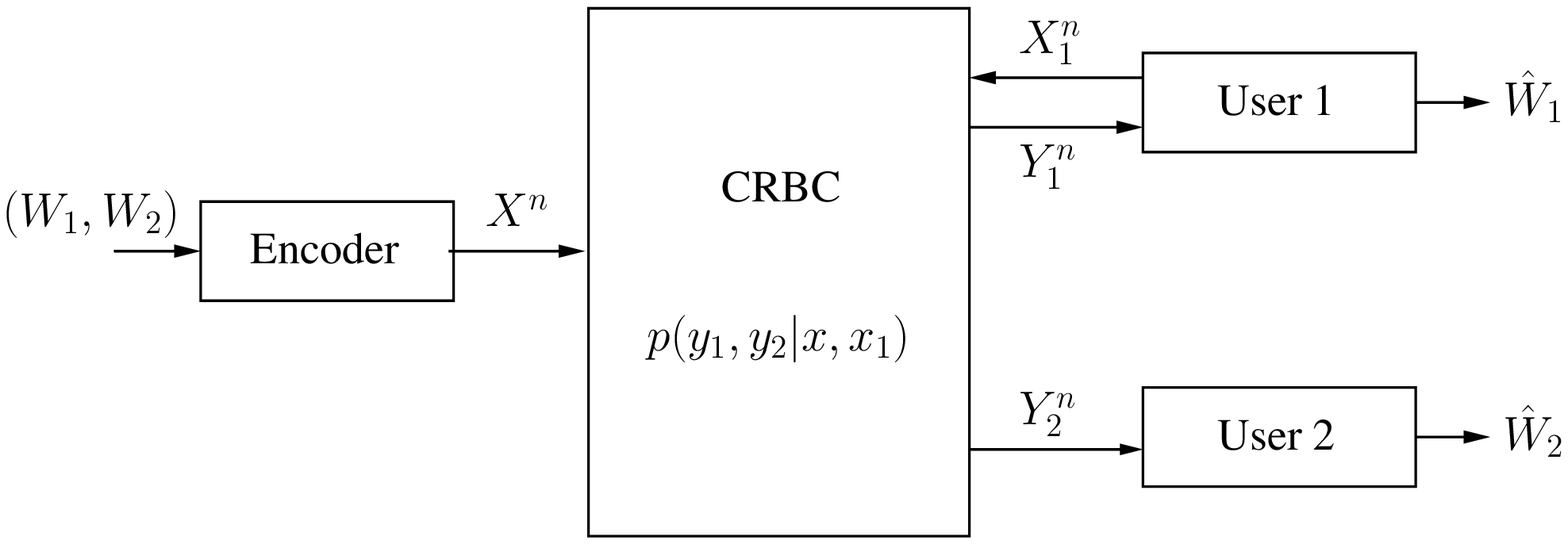}
\end{center}
\caption{Cooperative relay broadcast channel (CRBC) with single-sided
cooperative link.} \label{fig_channel}
\end{figure}

\begin{figure}[t]
\begin{center}
\epsfxsize = 12 cm  \epsffile{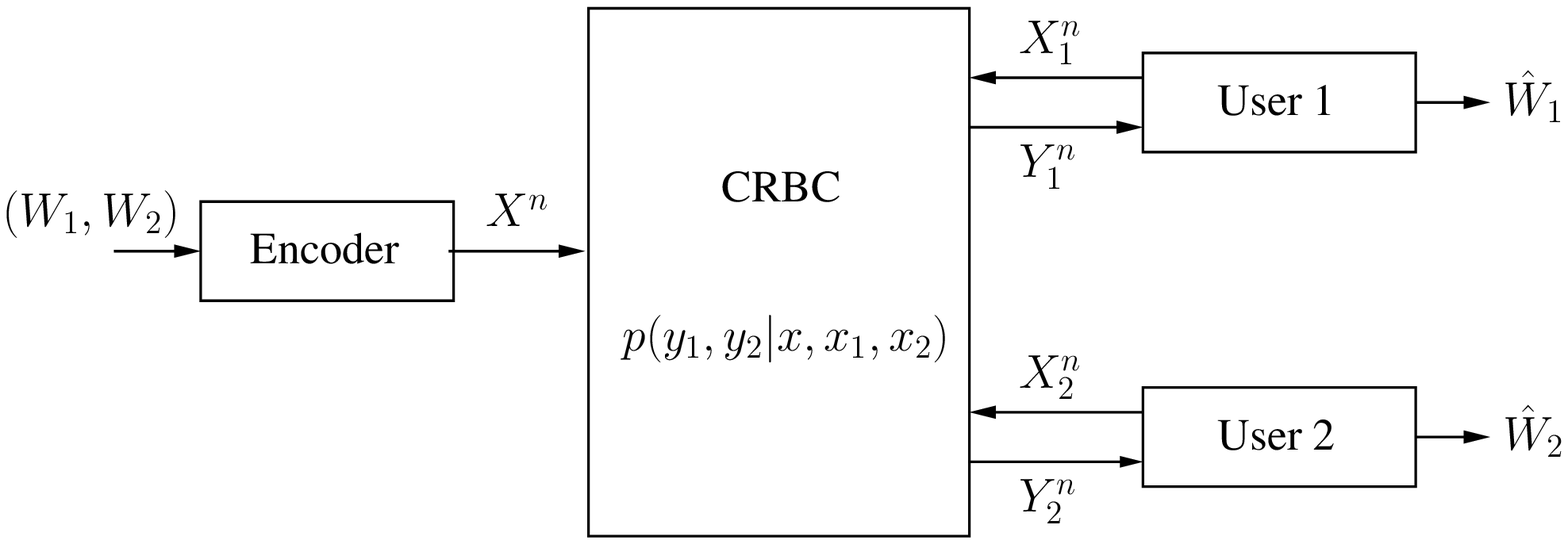}
\end{center}
\caption{Cooperative relay broadcast channel (CRBC) with a two-sided
cooperation link.} \label{fig_channel_two_sided}
\end{figure}

Our goal is to study the effects cooperation on the secrecy of
{\it multiple users} where secrecy refers to simultaneous individual
confidentiality of all users. One of the simplest models to study
this interaction is the cooperative relay broadcast channel
(CRBC), where there is a single transmitter and two receivers, and
each receiver would like to keep its message secret from the other
user; see Figures~\ref{fig_channel} and
\ref{fig_channel_two_sided}. In this model, in order to
incorporate the effects of cooperation, there is either a
single-sided (Figure~\ref{fig_channel}) or double-sided
(Figure~\ref{fig_channel_two_sided}) cooperative link between the
users. For clarity of ideas and simplicity of presentation, for a
major part of this paper, we will assume a CRBC with a single-sided
cooperation link from the first user to the second user. We will
investigate the effects of two-sided cooperation in
Section~\ref{sec:Two_sided}. Focusing on the single-sided CRBC, we
note that if we remove the cooperation link, our model reduces to
the BC with confidential messages in \cite{Ruoheng2, Ruoheng}, and
if we set the rate of the first user to zero, our model reduces to
the relay channel with confidential messages in \cite{Aylin_Yener,
Oohama,He_ISIT}, and if we both set the rate of the first user to
zero and remove the cooperation link between the users, our model
reduces to the single-user eavesdropper channel in \cite{Wyner,
Korner}. Our model is the simplest model (except perhaps for the
``dual'' model of cooperating transmitters in a MAC with per-user
secrecy constraints \cite{Ersen_Sennur}) that allows us to study
the effects of cooperation (or lack there of) of the first user
(the transmitting end of the cooperative link) on its own
equivocation rate as well as on the equivocation rate of the other
user (receiving end of the cooperative link).

Our motivation to study this problem can be best explained in a
Gaussian example. Imagine a two-user Gaussian BC. This BC is
degraded in one direction, hence both users cannot have positive
secrecy rates simultaneously \cite{Wyner,Ruoheng2, Ruoheng}. This
has motivated \cite{Ruoheng} to use multiple antennas at the
transmitter in order to remove this degradedness in either of the
directions and provide positive secrecy rates to both users
simultaneously. We wish to achieve a similar effect with a single
transmitter antenna, by introducing cooperation from one user to
the other. Imagine now a Gaussian CRBC \cite{Kramer,Servetto} as
in Figure~\ref{fig_channel}, where user 1 acts as a relay for user
2's message, i.e., that there is a cooperative link from user 1 to
user 2. Let us assume that in the underlying BC, user 1 has a
better channel. Without the cooperative link, user 2 cannot have
secure communication with the transmitter. We show that user 1 can
transmit cooperative signals and improve the secrecy rate of user
2. Our main
idea is that user 1 can use a compress-and-forward (CAF) based
relaying scheme for the message of user 2, and increase user 2's
rate to a level which is not decodable by user 1. This improves
user 2's secrecy. Now, let us assume that in the underlying BC,
user 1 has the worse channel. Without cooperation, user 1 cannot
have secure communication with the transmitter. We show that user
1 can transmit a jamming signal in the cooperative channel first
to guarantee a positive secrecy rate for itself assuming it has
enough power. This essentially brings the system to the setting
described in the previous case, and now user 1 can send a
cooperative signal to user 2 to help it achieve a positive secrecy
rate as well.

In this paper, we propose an achievable scheme that combines
Marton's coding scheme for BCs \cite{Marton} and Cover and El
Gamal's CAF scheme for relay channels \cite{Cover}. A similar
achievable scheme has appeared in \cite{Texas} which does not
consider any secrecy constraints, hence ours can be viewed as a
generalization of \cite{Texas} to a secrecy context. A similar achievable scheme also
appeared in \cite{Aylin_Yener,He_ISIT}, where CAF is applied to a
relay channel to provide improved secrecy for the main
transmitter. A relay channel can be considered as a special case
of the single-sided CRBC where the rate of the first user is set
to zero.

In this paper, we also develop a single-letter outer bound on the
rate-equivocation region; we accomplish singe-letterization by
using tools proposed in \cite{Korner}, namely by determining
suitable auxiliary random variables. Besides this outer bound, for
the second user, that is being helped in the single-sided CRBC, we develop another
single-letter outer bound which depends only on the channel inputs
and outputs.

To visualize the effects of cooperation on secrecy, we consider a
Gaussian CRBC and show that both users can have positive secrecy
rates through user cooperation. To obtain positive secrecy rates
for both users, we provide different assignments for the auxiliary
random variables appearing in the achievable rates. These
auxiliary random variable assignments have dirty paper coding
(DPC) interpretations \cite{costa}. In addition, we combine
jamming and relaying to provide secrecy for both users when the
relaying user is weak. Finally, we consider the CRBC with a
two-sided cooperation link and provide an achievable scheme for
this channel.

\section{The Channel Model and Definitions}
\label{sec:Channel_Model}

From here until the beginning of Section~\ref{sec:Two_sided}, we
will focus on a single-sided CRBC, and refer to it simply as CRBC.
The CRBC can be viewed as a relay channel where the transmitter
sends messages both to the relay node and the destination.
Therefore, one of the users, user 1 in our case, in a CRBC both
decodes its own message and also helps the other user. A CRBC
consists of two message sets $w_1\in \mathcal{W}_1, w_2\in
\mathcal{W}_2$, two input alphabets, one at the transmitter
$x\in\mathcal{X}$ and one at user 1 $x_1\in\mathcal{X}_1$, and two
output alphabets $y_1\in\mathcal{Y}_1, y_2\in\mathcal{Y}_2$, where
the former is for user 1 and the latter is for user 2. The channel
is assumed to be memoryless and its transition probability
distribution is $p(y_1,y_2|x,x_1)$.

A $\left(2^{nR_1},2^{nR_2},n\right)$ code for this channel
consists of two message sets as
$\mathcal{W}_1=\left\{1,\ldots,2^{nR_1}\right\}$ and
$\mathcal{W}_2=\left\{1,\ldots,2^{nR_2}\right\}$, an encoder at
the transmitter with mapping $\mathcal{W}_1\times
\mathcal{W}_2\rightarrow \mathcal{X}^{n}$, a set of relay
functions at user 1, $x_{1,i}=f_{i}(y_{1,1},\ldots,y_{1,i-1})$ for
$1\leq i\leq n$, two decoders, one at each user with the mappings
$g_1:\mathcal{Y}_1^{n}\rightarrow\mathcal{W}_1$ and
$g_2:\mathcal{Y}_2^{n}\rightarrow \mathcal{W}_2$. The probability
of error is defined as
$P_e^{n}=\max\left\{P_{e,1}^{n},P_{e,2}^{n}\right\}$ where
$P_{e,1}^{n}=\Pr\left(g_1(Y_1^n)\neq W_1 \right),
P_{e,2}^{n}=\Pr\left(g_2(Y_2^n)\neq W_2 \right)$. The secrecy of
the users is measured by the equivocation rates which are
$\frac{1}{n}H(W_1|Y_2^n) \textrm{ and } \frac{1}{n}H(W_2|Y_1^n,X_1^n)$.
Since user 1 has its own
channel input, we condition the entropy rate of user 2's messages on this
channel input.

A rate tuple $(R_1,R_2,R_{e,1},R_{e,2})$ is said to be achievable
if there exists a $\left(2^{nR_1},2^{nR_2},n\right)$ code with
$\lim_{n\rightarrow\infty} P_e^{n}=0$ and
\begin{eqnarray}
\lim_{n\rightarrow\infty} \frac{1}{n}H(W_1|Y_2^{n})\geq
R_{e,1},\quad \lim_{n\rightarrow\infty}
\frac{1}{n}H(W_2|Y_1^{n},X_1^n)\geq  R_{e,2}
\end{eqnarray}

\section{An Achievable Scheme} \label{sec:An achievable scheme}

We now provide an achievable scheme which combines Marton's coding
scheme for BCs \cite{Marton} and Cover and El Gamal's CAF scheme
for relay channels \cite{Cover}. A similar achievable scheme has
appeared in \cite{Texas} without any secrecy considerations. In
this scheme, user 1 sends a quantized version of its observation
to user 2, which uses this information to decode its own message.
The corresponding achievable rate-equivocation region is given by
the following theorem.

\begin{Theo}
\label{Theorem_1} The rate tuples $(R_1,R_2,R_{e,1},R_{e,2})$
satisfying
\begin{eqnarray}
R_1&\leq & I(V_1;Y_1|X_1) \label{theo1_R1}\\
R_2&\leq & I(V_2;Y_2,\hat{Y}_1|X_1) \label{theo1_R2}\\
R_1+R_2&\leq & I(V_1;Y_1|X_1)+I(V_2;Y_2,\hat{Y}_1|X_1)-I(V_1;V_2)
\label{theo1_R1_plus_R2}\\
R_{e,1}&\leq & R_1\\
R_{e,1}&\leq & \left[
I(V_1;Y_1|X_1)-I(V_1;Y_2,\hat{Y}_1|V_2,X_1)-I(V_1;V_2)\right]^+ \label{theo1_Re1}\\
R_{e,2} &\leq& R_2\\
R_{e,2} &\leq& \left[
I(V_2;Y_2,\hat{Y}_1|X_1)-I(V_2;Y_1|V_1,X_1)-I(V_1;V_2)\right]^+
\label{theo1_Re2}
\end{eqnarray}
are achievable for any distribution of the form
\begin{equation}
p(v_1,v_2)p(x|v_1,v_2)p(x_1)p(\hat{y}_1|x_1,v_1,y_1)p(y_1,y_2|x,x_1)
\label{Achievable_pdf}
\end{equation}
subject to the constraint
\begin{equation}
I(\hat{Y}_1;Y_1|X_1,V_1)\leq I(\hat{Y}_1,X_1;Y_2)
\label{Compression_constraint_gen}
\end{equation}
\end{Theo}
This theorem is a special case of
Theorem~\ref{Jamming_and_Relaying_Thm} and obtained from the
latter by setting $U=X_1$. Therefore, we will omit the proof of
Theorem~\ref{Theorem_1} here and will provide the proof of
Theorem~\ref{Jamming_and_Relaying_Thm} in
Appendix~\ref{Proof_of_Jamming_and_Relaying_Thm}. In
(\ref{theo1_Re1}) and (\ref{theo1_Re2}), $(x)^+$ is the positivity
operator, i.e., $(x)^+=\max(0,x)$.

\begin{Remark} We note that both the form of
the probability distribution in (\ref{Achievable_pdf}) and the
constraint in (\ref{Compression_constraint_gen}) in
Theorem~\ref{Theorem_1} are somewhat different than those of the
classical CAF scheme in \cite{Cover}. First, we condition the
distribution of $\hat{Y}_1$ on $V_1$ to prevent the compressed
version of $Y_1$ to leak any additional information regarding user
1's message on top of what user 2 already has through its own
observation. The constraint in (\ref{Compression_constraint_gen})
also reflects this concern. Similar constraints on the
distribution of $\hat{Y}_1$ and on the compression rate have
appeared in \cite{Texas}, where these modifications are not due to
secrecy constraints contrary to here. In \cite{Texas}, these are
imposed to obtain higher rates for user 2 by removing user 1's
private message from the compressed signal, whereas here, they are
imposed not to let $\hat{Y}_1$ leak any additional information
regarding user 1's message. Moreover, if we let user 1 compress
its observation without erasing its own message from the
observation, i.e., if we change the conditional distribution of
$\hat{Y}_1$ to $p(\hat{y}_1|x_1,y_1)$, we can recover the
constraint in \cite{Cover} (see equations (29)-(31) in
\cite{Texas}).
\end{Remark}

\begin{Remark}
\label{Remark_Broadcast_Ruoheng} If we disable the assistance of
user 1 to user 2 by setting $X_1=\hat{Y}_1=\phi$, the channel
model reduces to the BC with secrecy constraints, and the
achievable equivocation region becomes
\begin{eqnarray}
R_{e,1}^{BC}&\leq &
I(V_1;Y_1)-I(V_1;Y_2|V_2)-I(V_1;V_2) \label{BC_Re1} \\
R_{e,2}^{BC} &\leq& I(V_2;Y_2)-I(V_2;Y_1|V_1)-I(V_1;V_2) \label{BC_Re2}
\end{eqnarray}
where we require the Markov chain $(V_1,V_2)\rightarrow
X\rightarrow (Y_1,Y_2)$. This result was derived in
\cite{Ruoheng}.
\end{Remark}

\begin{Remark} If we disable both cooperation between receivers
by setting $X_1=\hat{Y}_1=\phi$, and also the confidential
messages sent to user 1 by setting $V_1=\phi$, the channel model
reduces to the single-user eavesdropper channel, and the
achievable equivocation rate for the second user becomes
\begin{equation}
R_{e,2}\leq I(V_2;Y_2)-I(V_2;Y_1)
\end{equation}
and the Markov chain $V_2\rightarrow X\rightarrow (Y_1,Y_2)$ is
required by the probability distribution in
(\ref{Achievable_pdf}). This is exactly the secrecy capacity of
the single-user eavesdropper channel given in \cite{Korner}.
\end{Remark}

\begin{Remark} If we disable the confidential messages sent to user
1 by setting $V_1=\phi$, the channel model reduces to a relay
channel with secrecy constraints, and the achievable equivocation
rate for the second user becomes
\begin{equation}
R_{e,2}\leq I(V_2;Y_2,\hat{Y}_1|X_1)-I(V_2;Y_1|X_1)
\end{equation}
subject to
\begin{equation}
I(\hat{Y}_1;Y_1|X_1)\leq I(\hat{Y}_1,X_1;Y_2)
\end{equation}
and the corresponding joint distribution reduces to
$p(v_2,x)p(x_1)p(\hat{y}_1|x_1,y_1)p(y_1,y_2|x,x_1)$. Further, if
we make the potentially suboptimal selection of $V_2=X$, the
corresponding achievable secrecy rate and the constraint coincide
with their counterparts found in \cite{Aylin_Yener} for the relay
channel.
\end{Remark}

\begin{Remark}
By comparing the equivocation rates of the users in
(\ref{theo1_Re1}) and (\ref{theo1_Re2}) and the equivocation rates
of the users in the corresponding BC given in
(\ref{BC_Re1}) and (\ref{BC_Re2}), we observe that the equivocation
rate of user 1 may decrease depending on the information contained
in $\hat{Y}_1$ and the equivocation rate of user 2 may increase
depending on the channel conditions.
\end{Remark}

\begin{Remark} We will show in the next section, where we develop outer
bounds for the rate-equivocation region, that if the channel of
user 2 is degraded with respect to the channel of user 1 then
$R_{e,2}=0$ (see Remark \ref{Remark_Outer_Bound}), where
degradedness is defined through the Markov chain $X\rightarrow
(X_1,Y_1)\rightarrow Y_2$. Here, we show, as an interesting
evaluation, that this achievable scheme cannot yield any positive
secrecy rates in this case, as expected.
\begin{align}
I(V_2;Y_2,\hat{Y}_1|X_1)&-I(V_2;Y_1|V_1,X_1)-I(V_1;V_2)\nonumber\\
&\leq I(V_2;Y_2,\hat{Y}_1,V_1|X_1)-I(V_2;Y_1|V_1,X_1)-I(V_1;V_2)\\
&=
I(V_2;Y_2,\hat{Y}_1|V_1,X_1)+I(V_2;V_1|X_1)-I(V_2;Y_1|V_1,X_1)-I(V_1;V_2)\\
&=
I(V_2;Y_2,\hat{Y}_1|V_1,X_1)-I(V_2;Y_1|V_1,X_1)\label{X1_V1V2_ind}\\
&\leq
I(V_2;Y_2,\hat{Y}_1,Y_1|V_1,X_1)-I(V_2;Y_1|V_1,X_1)\\
&=
I(V_2;Y_2,Y_1|V_1,X_1)+I(V_2;\hat{Y}_1|V_1,X_1,Y_1,Y_2)-I(V_2;Y_1|V_1,X_1)\\
&=
I(V_2;Y_2,Y_1|V_1,X_1)-I(V_2;Y_1|V_1,X_1)\label{hat_Y1_V1X1Y1_ind}\\
&= I(V_2;Y_2|V_1,X_1,Y_1)\\
&=0\label{itis_degraded}
\end{align}
where in (\ref{X1_V1V2_ind}), we used the fact that $X_1$ and
$(V_1,V_2)$ are independent, i.e., $I(V_1;V_2|X_1)=I(V_1;V_2)$, in
(\ref{hat_Y1_V1X1Y1_ind}), we used the Markov chain
$(V_2,Y_2)\rightarrow (V_1,X_1,Y_1)\rightarrow \hat{Y}_1$ which
implies $I(V_2;\hat{Y}_1|V_1,X_1,Y_1,Y_2)=0$, and in
(\ref{itis_degraded}), we used the Markov chain
$(V_1,V_2)\rightarrow X\rightarrow(X_1,Y_1)\rightarrow Y_2$ which
is due to the assumed degradedness.
\end{Remark}

\section{An Outer Bound}\label{sec:Outer_Bound}

We now provide an outer bound for the rate-equivocation region.
Our first outer bound in Theorem~\ref{Outer_Bound_Lemma} uses
auxiliary random variables. Next, in Theorem~\ref{cor_Re2}, we
provide a simpler outer bound for user 2 using only the channel
inputs and outputs, without employing any auxiliary random
variables.

\begin{Theo} \label{Outer_Bound_Lemma} The rate-equivocation region of the
CRBC lies in the union of the following rate
tuples\footnote{Unfortunately, in the conference
version~\cite{ISIT_08} of this paper, the outer bound appeared
with some typos.}
\begin{eqnarray}
R_1 &\leq & I(V_1;Y_1|X_1)\\
R_2 &\leq & I(V_2;Y_2)\\
R_{e,1}&\leq & \min \left\{ \tilde{R}_{e,1}, \bar{R}_{e,1}, R_1 \right\}\\
R_{e,2}&\leq & \min \left\{ \tilde{R}_{e,2}, \bar{R}_{e,2},
R_2\right\}
\end{eqnarray}
where
\begin{eqnarray}
\tilde{R}_{e,1} &=& I(V_1;Y_1|U)-I(V_1;Y_2|U)\\
\tilde{R}_{e,2} &=& I(V_2;Y_2|U)-I(V_2;Y_1|U)\\
\bar{R}_{e,1} &=& I(V_1;Y_1|V_2)-I(V_1;Y_2|V_2)\\
\bar{R}_{e,2} &=& I(V_2;Y_2|V_1)-I(V_2;Y_1|V_1)
\end{eqnarray}
where the union is taken over all joint distributions satisfying
the Markov chain
\begin{equation}
U\rightarrow\left(V_1,V_2\right)\rightarrow\left(X,X_1,Y_1\right)
\rightarrow Y_2 \label{Markov_Chain_Outer}
\end{equation}
\end{Theo}
The proof of this theorem is given in
Appendix~\ref{Proof_of_outer_bound}.

\begin{Remark} The bounds on the equivocation rates in
Theorem~\ref{Outer_Bound_Lemma} and those in \cite{Ruoheng}, where
the outer bounds are for the equivocation rates in a two-user BC
with per-user secrecy constraints as in here, have the same
expressions. The only difference between the two outer bounds is
in the Markov chain over which the union is taken. The Markov
chain in (\ref{Markov_Chain_Outer}) contains the one in
\cite{Ruoheng}, which is
\begin{equation}
U\rightarrow\left(V_1,V_2\right)\rightarrow X \rightarrow
\left(Y_1,Y_2\right)
\end{equation}
which means that our outer bound here evaluates to a larger region
than the one in \cite{Ruoheng}. This should be expected since the
achievable rate-equivocation region here in our CRBC contains the
achievable region in the BC.
\end{Remark}

We also provide a simpler outer bound for the equivocation
rate of user 2 which does not involve any auxiliary random
variables.

\begin{Theo}
\label{cor_Re2}
The equivocation
rate of user 2 is bounded as follows
\begin{equation}
R_{e,2} \leq \max_{p(x,x_1)}I(X;Y_2|X_1,Y_1) \label{simple_outer}
\end{equation}
\end{Theo}
The proof of this theorem is given in
Appendix~\ref{proof_simpler_outer_bound}.

\begin{Remark} \label{Remark_Outer_Bound} If the channel is
degraded, then the equivocation rate of user 2 is zero, since
\begin{align}
I(X;Y_2|X_1,Y_1)=0
\end{align}
which follows from the Markov chain
$X\rightarrow (X_1,Y_1)\rightarrow Y_2$ which
is a consequence of the degradedness.
\end{Remark}

\begin{Remark}
\label{Remark_Sato_Outer_Bound} We generally expect the outer
bound in Theorem~\ref{cor_Re2} to be loose because it
essentially assumes that user 2 has a complete access to user 1's
observation\footnote{In fact, this Sato-type \cite{Sato}
upper-bounding technique is used as a first step (before
introducing noise correlation to tighten the upper bound) in
finding the secrecy capacity of the MIMO wiretap channel
\cite{Hassibi,Ulukus,Wornell,Tie_Liu}.} whereas, in reality, user
2 has only limited information about user 1's observation, which
it obtains through the cooperative link. However, if the link from
user 1 to user 2 is strong enough, user 1 may be able to convey its
observation to user 2 precisely in which case the outer bound in
Theorem~\ref{cor_Re2} can be close to the achievable rate
obtained via the CAF scheme. For example, such a situation arises
if the channel satisfies the following Markov chain
\begin{align}
X\rightarrow (X_1,Y_2)\rightarrow Y_1\label{reverse_degraded}
\end{align}
For such channels, by selecting $V_2=X,V_1=\hat{Y}_1=\phi$ in the
achievable scheme, we get the following equivocation rate for user
2
\begin{align}
I(X;Y_2|X_1)-I(X;Y_1|X_1)=I(X;Y_2,Y_1|X_1)-I(X;Y_1|X_1)=I(X;Y_2|X_1,Y_1)
\end{align}
where the first equality is due to the Markov chain in
(\ref{reverse_degraded}). Hence, the outer bound in
(\ref{simple_outer}) gives the secrecy capacity for channels
satisfying (\ref{reverse_degraded}).
\end{Remark}
\begin{Remark}
Although we are able to provide a simple outer bound for the
equivocation rate of user 2, that depends only on the channel inputs
and outputs, finding such a simple outer bound for the
equivocation rate of user 1 does not seem to be possible. One reason
for this is that, user 1 can use its observation, i.e., $Y_1$, for
encoding its input, i.e., $X_1$, and create correlation between
its channel inputs and outputs across time. Consequently, this
correlation cannot be accounted for without using auxiliary random
variables. Another reason will be discussed in
Remark~\ref{jamming_perspective}.
\end{Remark}

\section{An Example: Gaussian CRBC}
\label{sec:Gaussian_Channels}

We now provide an example to show how the proposed achievable
scheme can enlarge the secrecy region for a Gaussian BC. The
channel outputs of a Gaussian CRBC are
\begin{align}
Y_1&=X+Z_1\\
Y_2&=X+X_1+Z_2
\end{align}
where $Z_1\sim\mathcal{N}(0,N_1), Z_2\sim\mathcal{N}(0,N_2)$ and are
independent, $E\left[X^2\right]\leq P, E\left[X_1^2\right]\leq
aP$. In this section, we assume that $N_2>N_1$, i.e., user 1 has a
stronger channel in the corresponding BC. Note that, in this case,
if user 1 does not help user 2, e.g., in the corresponding BC,
$R_{e,2}=0$. We present two different achievable schemes for this
channel where each one corresponds to a particular selection of
the underlying random variables in Theorem~\ref{Theorem_1}
satisfying the probability distribution condition in
(\ref{Achievable_pdf}). Proposition \ref{Independent_Inputs}
assigns independent channel inputs for each user, whereas
Proposition \ref{Dirty_paper_User1} uses a DPC scheme. For simplicity, we
provide only the achievable equivocation region in the following
propositions.

\begin{Prop}
\label{Independent_Inputs} The following equivocation rates are
achievable for all $\alpha \in[0,1]$
\begin{eqnarray}
R_{e,1}&\leq & \frac{1}{2}\log \left(1+\frac{\alpha P}{\bar{\alpha}P+N_1}
\right)-\frac{1}{2}\log \left(1+\frac{\alpha P}{N_2}\right)\\
R_{e,2}&\leq & \frac{1}{2}\log \left(1+\bar{\alpha}P\left(\frac{1}{\alpha P+
N_2}+\frac{1}{N_1+N_c}\right)\right)-\frac{1}{2}\log
\left(1+\frac{\bar{\alpha}P}{N_1}\right)
\end{eqnarray}
where $\bar{\alpha}=1-\alpha$ and $N_c$ is subject to
\begin{equation}
N_c\geq\frac{N_2(\bar{\alpha}P+N_1)+P(\alpha\bar{\alpha}P+N_1)}{aP}
\end{equation}
\end{Prop}

\begin{proof}This achievable region can be obtained by selecting
$V_1\sim \mathcal{N}(0,\alpha P),
V_2\sim\mathcal{N}(0,\bar{\alpha}P)$,
 $X=V_1+V_2$,
 $X_1\sim\mathcal{N}(0,aP)$,
$\hat{Y}_1=Y_1-V_1+Z_c=V_2+Z_1+Z_c$ and
$Z_c\sim\mathcal{N}(0,N_c)$, where $V_1,V_2,X_1\textrm{ and }Z_c$
are independent. The rates are found by direct calculation of the
expressions in Theorem~\ref{Theorem_1} using the above selection
of random variables.
\end{proof}

This achievable region can be enlarged by introducing correlation
between $V_1,V_2$. Since a joint encoding is performed at the
transmitter, one of the users' signals can be treated as a
non-causally known interference, and DPC~\cite{costa} can be used.
In the following proposition, the transmitter treats user 2's
signal as a non-causally known interference.

\begin{Prop}
\label{Dirty_paper_User1}  The following equivocation rates are
achievable for any $\gamma$ and all $\alpha \in [0,1]$
\begin{align}
R_{e,1}  &\leq  \frac{1}{2} \log
\left(1+\frac{(\bar{\alpha}\gamma+\alpha)^2
P}{(\alpha+\gamma^{2}\bar{\alpha})N_1+(\gamma-1)^2\alpha\bar{\alpha}P}\right)
-\frac{1}{2}\log\left(1+\frac{\alpha P}{N_2}\right)
-\frac{1}{2}\log\left(1+\gamma^2\frac{\bar{\alpha}}{\alpha}\right)\\
R_{e,2}& \leq   \frac{1}{2}\log
\left(1+\frac{\bar{\alpha}P(N_1+N_c)+\bar{\alpha}(1-\gamma)^2
P(\alpha P+N_2)}{(\alpha P+N_2)(N_1+N_c)}\right)
-\frac{1}{2}\log\left(1+\frac{\alpha\bar{\alpha}(\gamma-1)^2P}{(\alpha+
\gamma^2\bar{\alpha})N_1}\right)\nonumber\\
& \quad
-\frac{1}{2}\log\left(1+\gamma^2\frac{\bar{\alpha}}{\alpha}\right)
\end{align}
where $\bar{\alpha}=1-\alpha$ and $N_c$ is subject to
\begin{equation}
N_c \geq \frac{-\eta+\sqrt{\eta^2+4\theta \omega}}{2\theta}
\end{equation}
where
\begin{align}
\theta &= a(\alpha+\bar{\alpha}\gamma^2) P\\
\eta &= \left(\alpha +\gamma^2
\bar{\alpha}\right)P\big[aN_1+(1-\gamma)^2 \bar{\alpha}P
(a+\bar{\alpha})\big]-(P+N_2)\left[N_1(\alpha+\gamma^2\bar{\alpha})+\alpha\bar{\alpha}
(\gamma-1)^2P\right]\\
\omega & =  \left\{(P+N_2)\left[(1-\gamma)^2 \bar{\alpha}P+N_1
\right]-(1-\gamma)^2\bar{\alpha}^2 P^2 \right\}   \left\{ N_1
\left(\alpha+\gamma^2\bar{\alpha}\right)+P\alpha \bar{\alpha}
(\gamma-1)^2 \right\}
\end{align}
\end{Prop}

\begin{proof}
These equivocation rates are obtained by applying DPC for user 1.
Let the channel input of the transmitter be $X=U_1+U_2 $ where
$U_1\sim\mathcal{N}(0,\alpha P),
U_2\sim\mathcal{N}(0,\bar{\alpha}P)$ and are independent. The
auxiliary random variables are selected as $ V_2=U_2, V_1 =
U_1+\gamma U_2$, where for user 1, the signal of user 2 is treated
as non-casually known interference at the transmitter. The channel
output of user 1 is compressed as $ \hat{Y}_1 =
Y_1-V_1+Z_c=(1-\gamma)U_2+Z_1+Z_c $ where
$Z_c\sim\mathcal{N}(0,N_c)$ is the compression noise. The channel
input of user 1 is selected as $X_1\sim\mathcal{N}(0,aP)$. Here,
again, $U_1,U_2,Z_c \textrm{ and }X_1$ are all independent.  The
rates are then found by direct calculation of the expressions in
Theorem~\ref{Theorem_1} using the above selection of random
variables.
\end{proof}

\begin{figure}[p]
\begin{center}
\epsfxsize = 12.5 cm \epsffile{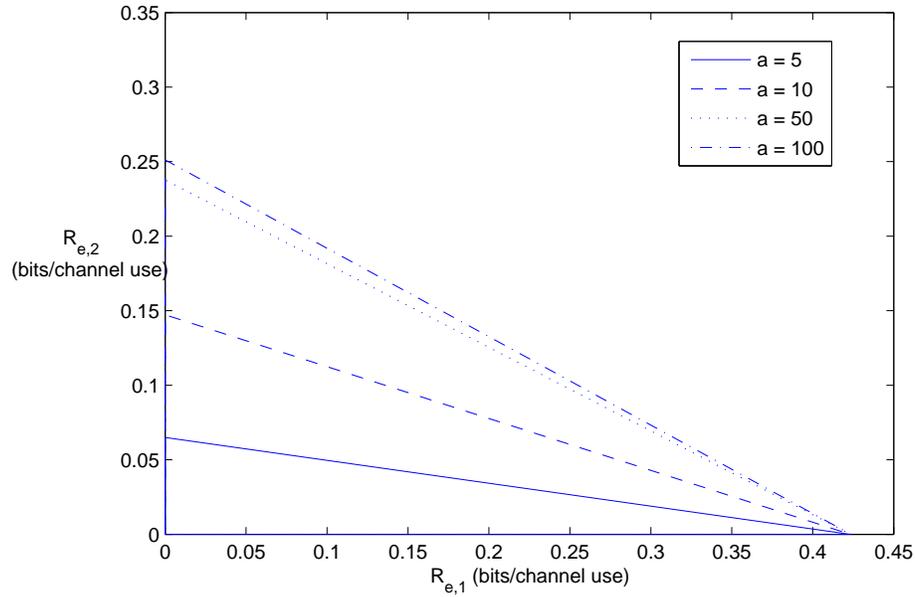}
\end{center}
\caption{Achievable equivocation rate region for single-sided CRBC
using Proposition~\ref{Independent_Inputs} where $V_1$ and $V_2$
are independent. $P=8,N_1=1,N_2=2$, i.e, user 2 has
no secrecy rate in the underlying BC.}
\label{Independent_Inputs_Fig}
\end{figure}
\begin{figure}[htp!]
\begin{center}
\epsfxsize = 12.5 cm \epsffile{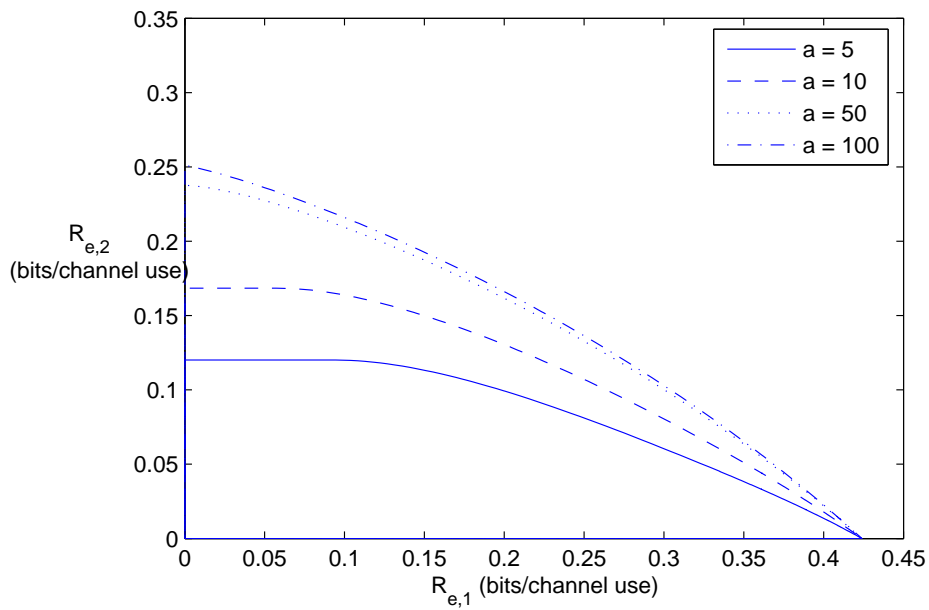}
\end{center}
\caption{Achievable equivocation region for single-sided CRBC
using Proposition~\ref{Dirty_paper_User1} where $V_1,V_2$ are
correlated, admitting a DPC interpretation. $P=8,N_1=1,N_2=2$,
i.e., user 2 has no secrecy rate in the underlying
BC.  } \label{Dirty_paper_User1_Fig}
\end{figure}

We note that, in both of the propositions above, $R_{e,2}$ is a
monotonically decreasing function of $N_c$. Consequently,
achievable $R_{e,2}$ depends on the quality of the cooperative
link between the users. If this link gets better allowing user 1
to convey its observation in a finer form, user 2's secrecy
increases. For illustrative purposes, the rate regions given by
Propositions~\ref{Independent_Inputs} and \ref{Dirty_paper_User1}
are evaluated for the parameters $P=8, N_1=1, N_2=2$, and the
corresponding plots are given in
Figures~\ref{Independent_Inputs_Fig} and \ref{Dirty_paper_User1_Fig}.
Note that since $N_2>N_1$, if there was no cooperation between the
users, user 2 could not have a positive secrecy rate. We observe
from these figures that, thanks to the cooperation of the users,
both users enjoy positive secrecy rates. However, we observe that
a positive secrecy for user 2 comes at the expense of a decrease
in the secrecy of user 1.

In particular, for both propositions, maximum secrecy rate for
user 2 is achieved when user 1 does not have any message itself
and acts as a pure relay for user 2. Similarly, user 1 achieves
the maximum secrecy rate when user 2 does not have any message.
Furthermore, we note that, for both achievable schemes, as
$a\rightarrow\infty$, the equivocation rate of user 2 approaches a
limit. This is due to the fact that, as $a\rightarrow\infty$, the
achievable equivocation rates are limited by the link between the
transmitter and user 1. Moreover, as $a\rightarrow\infty$, user 1
can send its observation to user 2 perfectly. Thus, in this case,
user 2 can be assumed to have a channel output of $(Y_1,Y_2)$,
which makes the channel of user 1 degraded with respect to the
channel of user 2. Consequently, following the analysis carried out
in Remark~\ref{Remark_Sato_Outer_Bound}, we expect the outer bound in
Theorem~\ref{cor_Re2} to become tight as $a\rightarrow\infty$, which is
stated in the next corollary.
\begin{Cor}
\label{Cor_a_infinity} As $a\rightarrow\infty$, the maximum
achievable equivocation rate for user 2 becomes
\begin{align}
R_{e,2}=\frac{1}{2}\log\left(1+P\left(\frac{1}{N_1}+\frac{1}{N_2}\right)\right)-\frac{1}{2}\log\left(1+\frac{P}{N_1}\right)
\label{max_secrecy_a_infinity}
\end{align}
\end{Cor}
The proof of this corollary is given in
Appendix~\ref{Proof_a_infinity}.

\section{Joint Jamming and Relaying}
\label{sec:jam_relay}
The proposed achievability scheme and its application to Gaussian
CRBC show us that user cooperation can enlarge the secrecy region.
However, this achievability scheme and the Gaussian example
provide us with only a limited picture of what can be achieved. In
particular, the achievability scheme proposed in
Section~\ref{sec:An achievable scheme} is designed with the
cooperating user (user 1) being the stronger of the two users in
mind. Next, we want to explore what can be done when the
cooperating user (user 1) is the weaker of the two users. In this
case, without the cooperative link, user 1 cannot have a positive
secrecy rate. Therefore, the first question to ask is, whether
user 1 can have a positive secrecy rate by utilizing the
cooperative link. The answer to this question is positive if user
1 uses the cooperative link to send a jamming signal to user 2.
However, a more interesting question is whether both users can
achieve positive secrecy simultaneously. The following theorem
provides an achievable scheme, where user 1 performs a combination
of jamming and relaying, to provide both users with positive
secrecy rates.

\begin{Theo}\label{Jamming_and_Relaying_Thm}
The rate quadruples $(R_1,R_2,R_{e,1},R_{e,2})$ satisfying
\begin{eqnarray}
R_1 &\leq & I(V_1;Y_1|X_1)\\
R_2 &\leq & I(V_2;Y_2,\hat{Y}_1|U)\\
R_1+R_2 &\leq &I(V_1;Y_1|X_1)+I(V_2;Y_2,\hat{Y}_1|U)-I(V_1;V_2)\\
R_{e,1}&\leq&R_1 \label{equi_user_1_jamming_relaying_2}\\
R_{e,1}&\leq & \left[
I(V_1;Y_1|X_1)-I(V_1;Y_2,\hat{Y}_1|V_2,U)-I(V_1;V_2)\right]^+\label{equi_user_1_jamming_relaying_1}\\
R_{e,2} &\leq& R_2\label{equi_user_2_jamming_relaying_2}\\
R_{e,2} &\leq &\left[
I(V_2;Y_2,\hat{Y}_1|U)-I(V_2;Y_1|V_1,X_1)-I(V_1;V_2)\right]^+
\label{equi_user_2_jamming_relaying_1}
\end{eqnarray}
are achievable for any distribution of the form
\begin{equation}
p(v_1,v_2)p(x|v_1,v_2)p(u)p(x_1|u)p(\hat{y}_1|u,v_1,y_1)p(y_1,y_2|x,x_1)
\label{distribution_jam_selfrelay}
\end{equation}
subject to the following constraint
\begin{equation}
I(\hat{Y}_1;Y_1|X_1,V_1,U)\leq I(\hat{Y}_1,U;Y_2)
\label{compression_constraint_jam_relay}
\end{equation}
\end{Theo}
The proof of this theorem is given in
Appendix~\ref{Proof_of_Jamming_and_Relaying_Thm}.

\begin{Remark} In Theorem~\ref{Jamming_and_Relaying_Thm}, $U$ denotes
the actual help signal, while the channel input $X_1$, which is
correlated with $U$, may include an additional jamming attack. The
intuition behind this achievable scheme is that, although user 2
should be able to decode $U$, it cannot decode the entire $X_1$.
Therefore, since user 2 cannot decode and eliminate $X_1$ from
$Y_2$, its channel becomes an attacked one, where decoding
$V_1$ may be impossible. Therefore, in this scheme, user 1
first attacks user 2 to make its channel worse by associating $U$
with many $X_1$s (hence, it confuses user 2), and then helps it to
improve its secrecy rate.
\end{Remark}

\begin{Remark} We note that this achievable scheme
is reminiscent of ``cooperative jamming'' \cite{Aylin_2}. In
\cite{Aylin_2}, the focus is on a two user MAC with an external
eavesdropper, where one of the users attacks both the legitimate
receiver and the eavesdropper, with the hope that it hurts the
eavesdropper more than it hurts the legitimate receiver, and
improves the secrecy of the legitimate receiver. In contrast, in
our work, the relay (user 1) attacks user 2 to improve its own
secrecy.
\end{Remark}

\section{Gaussian Example Revisited}
\label{sec:Gaussian_Revisited}
Consider again the Gaussian
CRBC, now with $N_1>N_2$. The scheme proposed in Theorem~\ref{Jamming_and_Relaying_Thm}
works as follows:
user 1 divides $X_1$ into two parts. The first part carries the
noise and the second part carries the bin index of $\hat{Y}_1$.
Although Theorem~\ref{Jamming_and_Relaying_Thm} is valid for all
cases, assume here that user 1 has large enough power. Then, the
first part makes user 2's channel noisier than user 1's channel.
This brings the situation to the case studied in
Section~\ref{sec:Gaussian_Channels}. Consequently, we can now have
a positive secrecy rate for user 1, and also provide a positive
secrecy rate to user 2, by sending a compressed version of $Y_1$
to it, as in Section~\ref{sec:Gaussian_Channels}.

\begin{Prop} \label{Prop_SelfRelaying_Independent_Inputs}
The following equivocation rates are achievable for all $
(\alpha,\beta)\in [0,1]\times [0,1]$
\begin{eqnarray}
R_{e,1}&\leq &\frac{1}{2} \log \left(1+\frac{\alpha
P}{\bar{\alpha}P+N_1} \right)-\frac{1}{2}\log \left(1+\frac{\alpha
P}{a\bar{\beta}P+N_2}\right)\\
R_{e,2} &\leq & \frac{1}{2}\log \left(1+\bar{\alpha}P\left(\frac{1}{N_1+N_c}
+\frac{1}{\alpha P+N_2+a\bar{\beta}P}\right)\right)-\frac{1}{2} \log \left
(1+\frac{\bar{\alpha}P}{N_1}\right)
\end{eqnarray}
where $\bar{\alpha}=1-\alpha,\bar{\beta}=1-\beta$, and $N_c$ is
subject to
\begin{equation}
N_c\geq \frac{\bar{\alpha}P(\alpha P +N_2+a\bar{\beta}P)+N_1
(P+N_2+a\bar{\beta}P)}{a\beta P}
\end{equation}
\end{Prop}

\begin{proof} This achievable region is obtained by selecting the random variables in
Theorem~\ref{Jamming_and_Relaying_Thm} as $X=V_1+V_2$ where
$V_1\sim\mathcal{N}(0,\alpha P),
V_2\sim\mathcal{N}(0,\bar{\alpha}P)$, $X_1=U+Z_j$ where
$U\sim\mathcal{N}(0,a\beta P),
Z_j\sim\mathcal{N}(0,a\bar{\beta}P)$,
$\hat{Y}_1=Y_1-V_1+Z_c=V_2+Z_1+Z_c$ where
$Z_c\sim\mathcal{N}(0,N_c)$. Moreover, $V_1,V_2,U,Z_j,Z_c$ are all
independent. Here, $Z_j$ serves as the jamming signal, and $U$
serves as the helper signal.  User 1 first jams user 2 and makes its
channel noisier than its own by using $Z_j$ and then helps user 2
through sending a compressed version of its observation by using
$U$. The rates are then found by direct calculation of the
expressions in Theorem~\ref{Jamming_and_Relaying_Thm} using the
above selection of random variables.
\end{proof}

Moreover, as in Section~\ref{sec:Gaussian_Channels}, we can use
DPC based schemes in this case also. The following proposition
characterizes the DPC scheme for
Theorem~\ref{Jamming_and_Relaying_Thm}.

\begin{Prop}
\label{Prop_SelfRelaying_DPC_User1}  The following equivocation
rates are achievable for any $\gamma$ and for all $ (\alpha,\beta)
\in [0,1]\times[0,1]$
\begin{align}
R_{e,1}  &\leq  \frac{1}{2} \log
\left(1+\frac{(\bar{\alpha}\gamma+\alpha)^2
P}{(\alpha+\gamma^{2}\bar{\alpha})N_1+(\gamma-1)^2\alpha\bar{\alpha}P}\right)
-\frac{1}{2}\log\left(1+\frac{\alpha
P}{(a\bar{\beta}P+N_2)}\right)\nonumber\\
&\quad
-\frac{1}{2}\log\left(1+\gamma^2\frac{\bar{\alpha}}{\alpha}\right)\\
R_{e,2}& \leq   \frac{1}{2}\log
\left(1+\frac{\bar{\alpha}P(N_1+N_c)+\bar{\alpha}(1-\gamma)^2
P(\alpha P+a\bar{\beta}P+N_2)}{(\alpha
P+a\bar{\beta}P+N_2)(N_1+N_c)}\right)\nonumber\\
&\quad
-\frac{1}{2}\log\left(1+\frac{\alpha\bar{\alpha}(\gamma-1)^2P}{(\alpha+
\gamma^2\bar{\alpha})N_1}\right)
-\frac{1}{2}\log\left(1+\gamma^2\frac{\bar{\alpha}}{\alpha}\right)
\end{align}
where $\bar{\alpha}=1-\alpha,\bar{\beta}=1-\beta$ and $N_c$ is
subject to
\begin{equation}
N_c \geq \frac{-\eta+\sqrt{\eta^2+4\theta \omega}}{2\theta}
\end{equation}
where
\begin{align}
\theta &= a\beta(\alpha+\bar{\alpha}\gamma^2) P\\
\eta &= \left(\alpha +\gamma^2 \bar{\alpha}\right)P\big[a\beta
N_1+(1-\gamma)^2 \bar{\alpha}P (a\beta+\bar{\alpha})\big]
\nonumber
\\
&\quad
-(P+a\bar{\beta}P+N_2)\left[N_1(\alpha+\gamma^2\bar{\alpha})+\alpha\bar{\alpha}
(\gamma-1)^2P\right]\\
\omega & = \left[(P+a\bar{\beta}+N_2)\left[(1-\gamma)^2
\bar{\alpha}P+N_1 \right]-(1-\gamma)^2\bar{\alpha}^2 P^2 \right]
\left[ N_1 \left(\alpha+\gamma^2\bar{\alpha}\right)+P\alpha
\bar{\alpha} (\gamma-1)^2 \right]
\end{align}
\end{Prop}

\begin{proof} All random variable selections are the same as in
Proposition~\ref{Dirty_paper_User1} except for $X_1,U$. Here, we
choose $X_1=Z_j+U$ and $U\sim\mathcal{N}(0,a\beta P),
Z_j\sim\mathcal{N}(0,a \bar{\beta}P)$. $U,Z_j$ are independent.
\end{proof}

\begin{figure}[p]
\begin{center}
\epsfxsize = 12.5 cm
\epsffile{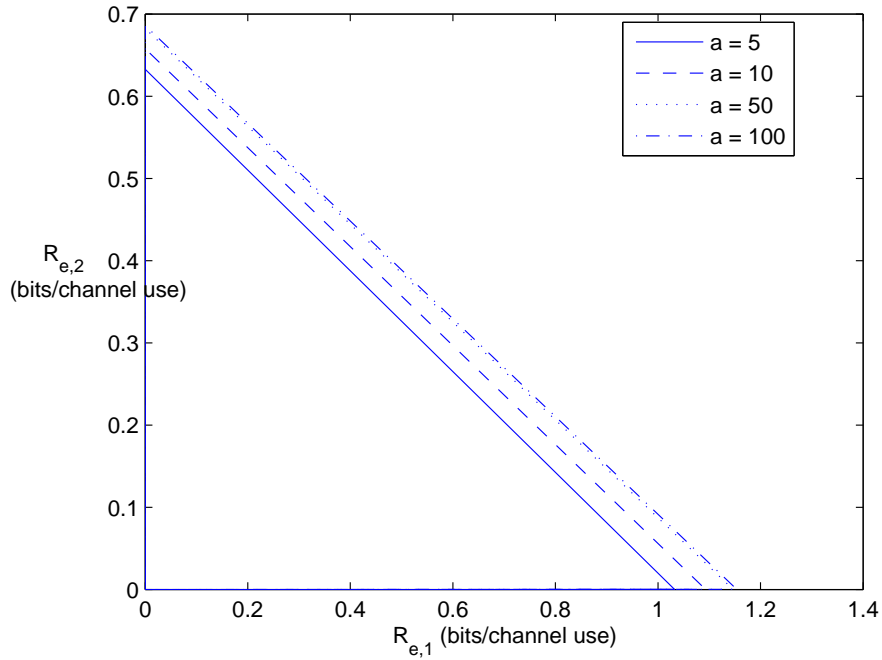}
\end{center}
\caption{Achievable equivocation rate region using
Proposition~\ref{Prop_SelfRelaying_Independent_Inputs} where user
1 jams and relays, and $V_1,V_2$ are independent.
$P=8,N_1=2,N_2=1$, i.e., user 1 cannot have any
positive secrecy in the underlying BC.}
\label{Jamming_Independent_Inputs_Fig}
\end{figure}

\begin{figure}[htp!]
\begin{center}
\epsfxsize = 12.5 cm
\epsffile{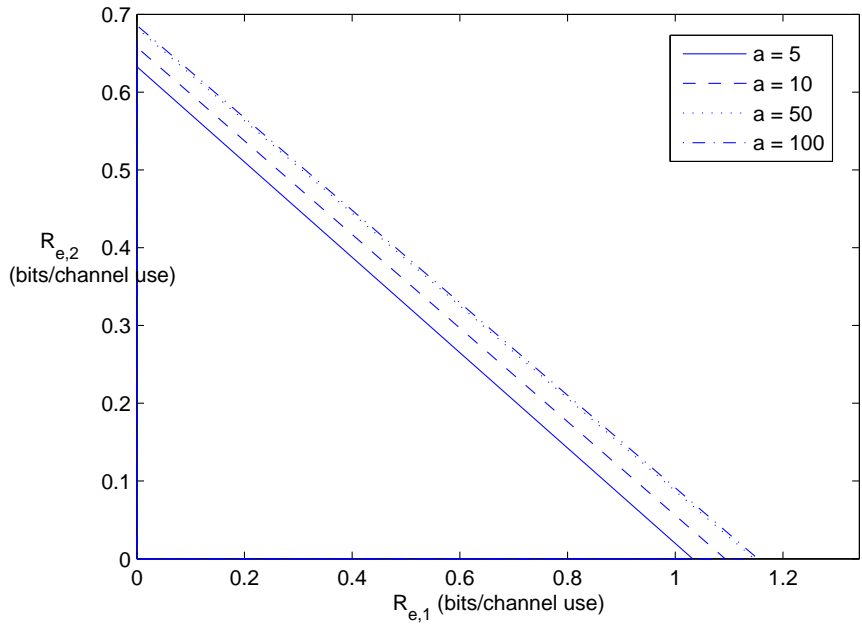}
\end{center}
\caption{Achievable equivocation rate region using
Proposition~\ref{Prop_SelfRelaying_DPC_User1} where user 1 jams
and relays, and $V_1,V_2$ are correlated, admitting a DPC
interpretation. $P=8,N_1=2,N_2=1$, i.e., user 1 cannot
have any positive secrecy in the underlying BC.}
\label{DPC_User1_Jam_Relay_Fig}
\end{figure}

\begin{figure}[p]
\begin{center}
\epsfxsize = 12.5 cm
\epsffile{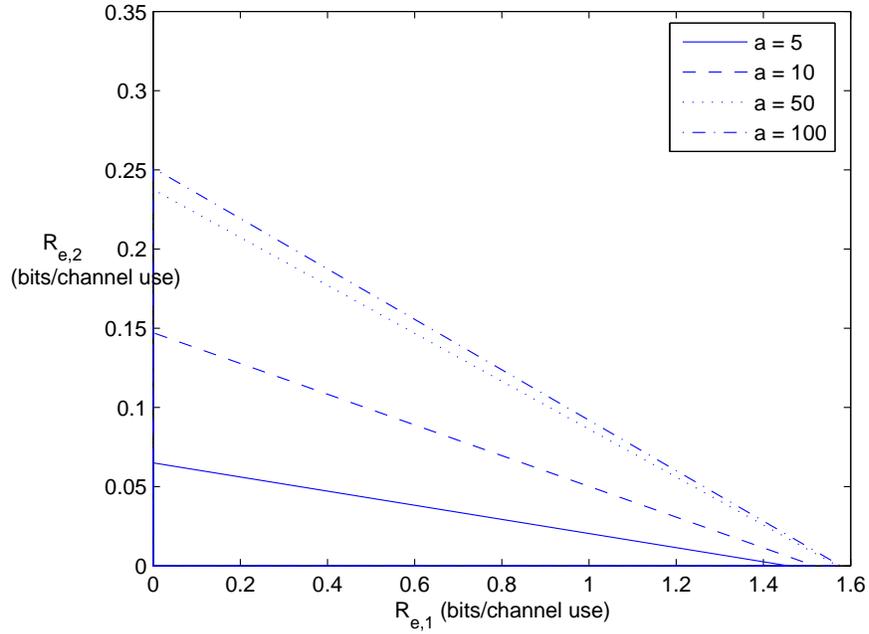}
\end{center}
\caption{Achievable equivocation rate region using
Proposition~\ref{Prop_SelfRelaying_Independent_Inputs} where user
1 jams and relays, and
$V_1,V_2$ are independent. $P=8,N_1=1,N_2=2$, i.e.,
user 1's channel is stronger than user 2.}
\label{Jam_Relay_Ind_Input_User1_Stronger_fig}
\end{figure}

\begin{figure}[htp!]
\begin{center}
\epsfxsize = 12.5 cm
\epsffile{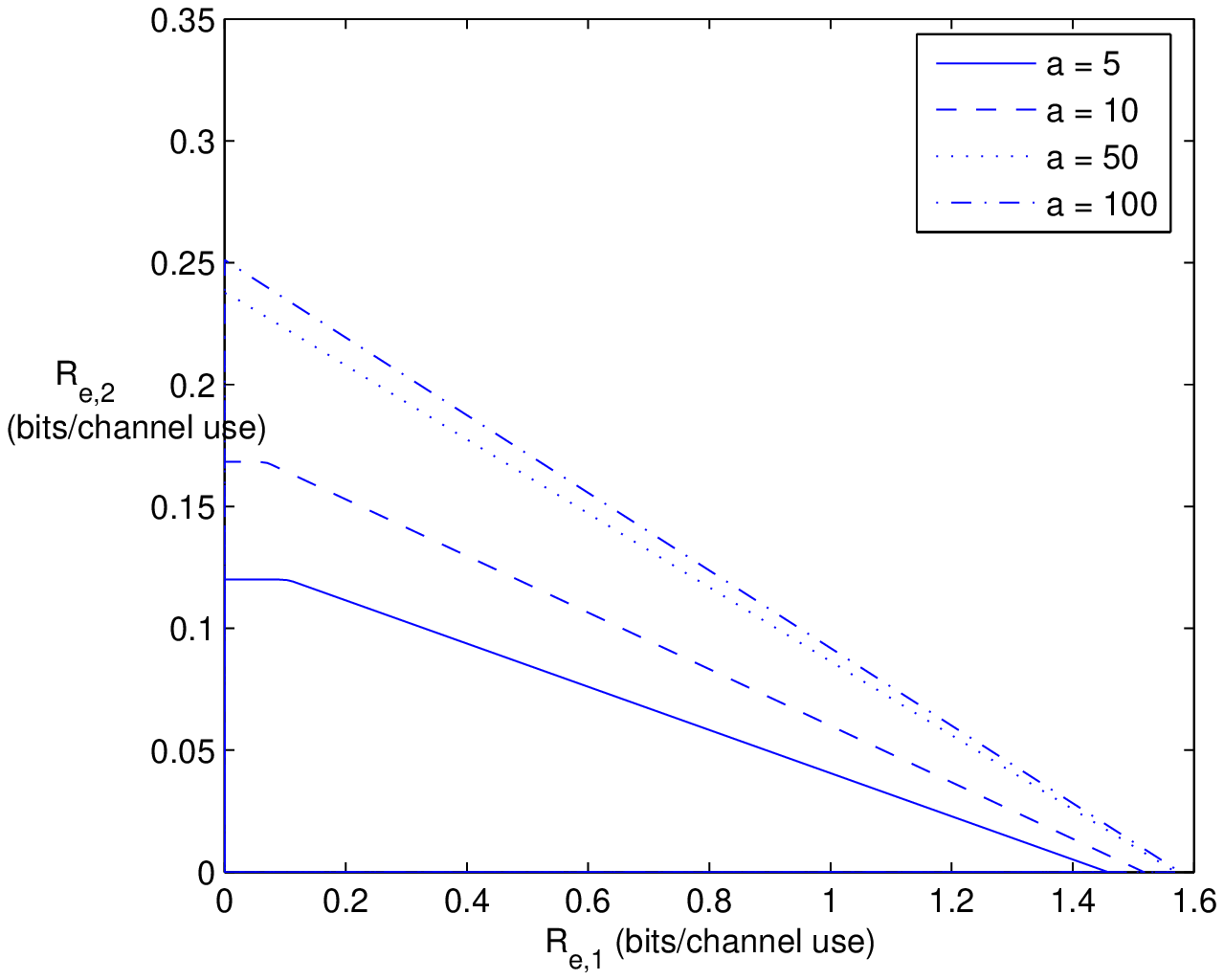}
\end{center}
\caption{Achievable equivocation rate region using
Proposition~\ref{Prop_SelfRelaying_DPC_User1} where user 1 jams and relays,
and $V_1,V_2$ are
correlated, admitting a DPC interpretation. $P=8, N_1=1, N_2=2$,
i.e., user 1's channel is stronger than user
2.} \label{DPC_User1_Jam_Relay_User1_Stronger_fig}
\end{figure}

We first note that
Propositions~\ref{Prop_SelfRelaying_Independent_Inputs},~\ref{Prop_SelfRelaying_DPC_User1}
reduce to
Propositions~\ref{Independent_Inputs},~\ref{Dirty_paper_User1},
respectively, by simply selecting $\beta=0$, i.e., no jamming. We
provide a numerical example in
Figures~\ref{Jamming_Independent_Inputs_Fig},~\ref{DPC_User1_Jam_Relay_Fig}
for $P=8, N_1=2, N_2=1$. Since $N_1>N_2$, a positive secrecy rate
for user 1 would not be possible if the cooperative link did not
exist. However, if user 1 has enough power to make user 2's
channel noisier by injecting Gaussian noise to it, user 1 can
provide secrecy for itself. For user 1 to have positive secrecy, we need
\begin{align}
a\geq \frac{N_1-N_2}{P}
\end{align}
Otherwise, user 1 cannot have positive secrecy by using strategies
employed in
Propositions~\ref{Prop_SelfRelaying_Independent_Inputs},
\ref{Prop_SelfRelaying_DPC_User1}. In addition, contrary to
Section~\ref{sec:Gaussian_Channels}, we observe from
Figures~\ref{Jamming_Independent_Inputs_Fig}
and~\ref{DPC_User1_Jam_Relay_Fig} that here DPC based schemes do
not provide any gain with respect to the independent selection of
$V_1,V_2$. Furthermore, we also apply
Propositions~\ref{Prop_SelfRelaying_Independent_Inputs} and
\ref{Prop_SelfRelaying_DPC_User1} to the case where user 1 is
stronger than user 2 by selecting the noise variances as
$N_1=1,N_2=2$ as in Section~\ref{sec:Gaussian_Channels} to show
that propositions presented in this section cover the ones in Section~\ref{sec:Gaussian_Channels}.
We provide the corresponding graphs in
Figures~\ref{Jam_Relay_Ind_Input_User1_Stronger_fig}
and~\ref{DPC_User1_Jam_Relay_User1_Stronger_fig}. Comparing
Figures~\ref{Independent_Inputs_Fig}
(resp.~\ref{Dirty_paper_User1_Fig})
and~\ref{Jam_Relay_Ind_Input_User1_Stronger_fig}
(resp.~\ref{DPC_User1_Jam_Relay_User1_Stronger_fig}), we observe
that even though the maximum secrecy rate of user 2 remains the
same, the maximum secrecy rate of user 1 is improved
significantly. This improvement comes, because through
Propositions~\ref{Prop_SelfRelaying_Independent_Inputs}
and~\ref{Prop_SelfRelaying_DPC_User1}, user 1 jams the receiver of
user 2.



Next, we examine Figures~\ref{Independent_Inputs_Fig}
and~\ref{Jam_Relay_Ind_Input_User1_Stronger_fig} in more detail.
In Figure~\ref{Independent_Inputs_Fig}, for instance when $a=100$,
the largest $R_{e,2}$, which is about 0.25 bits/channel use, is
obtained when $R_{e,1}=0$. This corresponds to the case where user
1's rate and secrecy rate are set to zero. In this case, user 1
serves as a pure relay for user 2. The secrecy rate we obtain at
this extreme is the same as \cite{Aylin_Yener,He_ISIT}. At the
other extreme, the largest $R_{e,1}$, which is about 0.42
bits/channel use, is obtained when $R_{e,2}=0$. In this case, user
2 is just an eavesdropper in a single-user channel from the
transmitter to user 1. The secrecy rate we obtain at this extreme
is the same as \cite{Wyner,Korner,hellman}. Moreover, as we
see from Figure~\ref{Independent_Inputs_Fig}, whenever user 1 helps
user 2 to have positive secrecy, it needs to deviate from this
extreme point. Thus, user 2's positive secrecy rates come at the
expense of a decrease in user 1's secrecy rate. If we consider
Figure~\ref{Jam_Relay_Ind_Input_User1_Stronger_fig}, the largest
$R_{e,2}$ is the same as that in
Figure~\ref{Independent_Inputs_Fig}, which is again achieved when
$R_{e,1}=0$, i.e., when user 1 acts as a pure relay for user 2.
However, in Figure~\ref{Jam_Relay_Ind_Input_User1_Stronger_fig},
user 1's maximum secrecy rate increases dramatically due to its
jamming capabilities in
Proposition~\ref{Prop_SelfRelaying_Independent_Inputs}. In
Figure~\ref{Jam_Relay_Ind_Input_User1_Stronger_fig}, user 1
achieves its maximum secrecy rate, which is about 1.58
bits/channel use, when it uses all of its power for jamming
user 2's receiver and when the rate of user 2 is set to zero. We note that
this rate is larger than that is achievable in the corresponding
single-user eavesdropper channel from the transmitter to user 1,
while user 2 is an eavesdropper. We
observe from Figure~~\ref{Jam_Relay_Ind_Input_User1_Stronger_fig} that when user 1 is able to jam and relay
jointly, it can provide secrecy for user 2 while its own secrecy
rate is still larger than that of the corresponding single-user
eavesdropper channel.
Thus, as opposed to the case where it can only relay, i.e.,
Proposition~\ref{Independent_Inputs}, both users enjoy secrecy in
Proposition~\ref{Prop_SelfRelaying_Independent_Inputs}, while user
1 does not have to compromise from its own secrecy rate that is achievable
in the underlying eavesdropper channel.

\begin{Remark}
\label{jamming_perspective} We are now ready to discuss why we
could not find an outer bound for the equivocation rate of user 1
that relies only on the channel inputs and outputs. To understand
this, we first examine the outer bound we found on the
equivocation rate of user 2 in Theorem~\ref{cor_Re2}. This outer
bound is obtained by giving the entire observation of user 1 to user
2 (i.e., $N_c=0$). Hence, this is the best possible scenario as
far as the channel of user 2 is concerned, and thus, it yields an
outer bound. However, a similar approach cannot work for user 1,
because although user 1 can have access to the observation of user
2, user 1 still has additional freedom (and opportunities) to
increase its own secrecy rate by sending jamming signals over the
cooperative link, as shown in this section. This is the main
reason why we could not find a simple outer bound for user 1's
secrecy rate using only the channel inputs/outputs.
\end{Remark}

\section{Two-sided Cooperation}
\label{sec:Two_sided} In this section, we provide an achievable
scheme for CRBC with two-sided cooperation. In this case, each
user can act as a relay for the other one; see
Figure~\ref{fig_channel_two_sided}. The corresponding channel
consists of two message sets $w_1\in \mathcal{W}_1, w_2\in
\mathcal{W}_2$, three input alphabets, one at the transmitter
$x\in\mathcal{X}$, one at user 1 $x_1\in\mathcal{X}_1$ and one at
user 2 $x_2\in\mathcal{X}_2$. The channel consists of two output
alphabets denoted by $y_1\in\mathcal{Y}_1, y_2\in\mathcal{Y}_2$ at
the two users. The channel is assumed to be memoryless and its
transition probability distribution is $p(y_1,y_2|x,x_1,x_2)$.

A $\left(2^{nR_1},2^{nR_2},n\right)$ code for this channel
consists of two message set as
$\mathcal{W}_1=\left\{1,\ldots,2^{nR_1}\right\}$ and
$\mathcal{W}_2=\left\{1,\ldots,2^{nR_2}\right\}$, an encoder at
the transmitter which maps each pair $(w_1,w_2)\in
\left(\mathcal{W}_1\times \mathcal{W}_2\right)$ to a codeword
$x^{n}\in\mathcal{X}^{n}$, a set of relay functions at user 1,
$x_{1,i}=f_{1,i}(y_{1,1},\ldots,y_{1,i-1}),\hspace{0.1cm}\newline 1\leq
i\leq n,$ and a set of relay functions at user 2,
$x_{2,i}=f_{2,i}(y_{2,1},\ldots,y_{2,i-1}),\hspace{0.1cm} 1\leq
i\leq n, $ two decoders, one at user 1 and one at user 2 with the
mappings $
g_1:\mathcal{Y}_1^{n}\rightarrow\mathcal{W}_1,\hspace{0.1cm}
g_2:\mathcal{Y}_2^{n}\rightarrow \mathcal{W}_2.$ Definitions for
the error probability for this two-sided case are the same as in the
single-sided case. The secrecy of the users is again measured
by the equivocation rates which are $\frac{1}{n}H(W_1|Y_2^n,X_2^n)$
and $\frac{1}{n}H(W_2|Y_1^n,X_1^n)$. In this case, since user 2 has
a channel input also, we condition the entropy rate of user 1's messages
on this channel input.

A rate tuple $(R_1,R_2,R_{e,1},R_{e,2})$ is said to be achievable
if there exists a $\left(2^{nR_1},2^{nR_2},n\right)$ code with
$\lim_{n\rightarrow\infty}P_e^n=0,$ and
\begin{align}
\lim_{n\rightarrow\infty}\frac{1}{n}H(W_1|Y_2^n,X_2^n)\geq R_{e,1},\quad
\lim_{n\rightarrow\infty}\frac{1}{n}H(W_2|Y_1^n,X_1^n)\geq R_{e,2}
\end{align}

The following theorem characterizes an achievable region for this channel model.
\begin{Theo}
\label{Two_sided} The rate tuples $(R_1,R_2,R_{e,1},R_{e,2})$
satisfying
\begin{eqnarray}
R_1&\leq& I(V_1;Y_1,\hat{Y}_2|X_1,U_2)  \\
R_2&\leq& I(V_2;Y_2,\hat{Y}_1|X_2,U_1)  \\
R_1+R_2 &\leq &
I(V_1;Y_1,\hat{Y}_2|X_1,U_2)+I(V_2;Y_2,\hat{Y}_1|X_2,U_1)-I(V_1;V_2)
\\
R_{e,1}&\leq & R_1\\
R_{e,1}&\leq & \left[
I(V_1;Y_1,\hat{Y}_2|X_1,U_2)-I(V_1;Y_2,\hat{Y}_1|V_2,X_2,U_1)-I(V_1;V_2)\right]^+
\\
R_{e,2}&\leq & R_2\\
R_{e,2}&\leq & \left[
I(V_2;Y_2,\hat{Y}_1|X_2,U_1)-I(V_2;Y_1,\hat{Y}_2|V_1,X_1,U_2)-I(V_1;V_2)\right]^+
\end{eqnarray}
are achievable for any distribution of the form
\begin{eqnarray}
p(v_1,v_2)p(x|v_1,v_2)p(u_1,x_1)p(\hat{y}_1|u_1,y_1)p(u_2,x_2)p(\hat{y}_2|u_2,y_2)p(y_1,y_2|x,x_1,x_2)
\label{achievable_pdf_two_sided}
\end{eqnarray}
subject to the following constraints
\begin{eqnarray}
I(\hat{Y}_1;Y_1|U_1,X_1,U_2)&\leq & I(\hat{Y}_1,U_1;Y_2|X_2)
\\
I(\hat{Y}_2;Y_2|U_2,X_2,U_1)&\leq & I(\hat{Y}_2,U_2;Y_1|X_1)
\end{eqnarray}
\end{Theo}
The proof of this theorem is given in
Appendix~\ref{Proof_of_two_sided}.

Contrary to the previous achievable schemes given in
Theorem~\ref{Theorem_1} and \ref{Jamming_and_Relaying_Thm}, here
users do not compress their observations after erasing their
codewords from the observations; this is why we did not condition
$\hat{Y}_1$ (resp. $\hat{Y}_2$) on $V_1$ (resp. $V_2$) in
(\ref{achievable_pdf_two_sided}). In fact, they cannot remove
their own codewords from their observations because each user
employs a sliding-window type decoding scheme, i.e., they should wait
until the next block to decode their own codewords, whereas
compression should be performed right after the reception of the
previous block, at which time they have not yet decoded their own
messages. However, we note that this achievable scheme also
provides opportunities for jamming as did the achievable scheme
provided in Section~\ref{sec:jam_relay}.

\section{Gaussian Example for Two-sided Cooperation}
The channel outputs of a
Gaussian CRBC with two-sided cooperation are
\begin{align}
Y_1&=X+X_2+Z_1\\
Y_2&=X+X_1+Z_2
\end{align}
where $Z_1\sim\mathcal{N}(0,N_1),Z_2\sim\mathcal{N}(0,N_2)$ and are
independent, $E\left[X^2\right]\leq P$, $E\left[X_1^2\right]\leq
a_1P$, $E\left[X_2^2\right]\leq a_2P$.

We present the following proposition which
characterizes an achievable equivocation region.

\begin{Prop}
\label{Prop_TwoSided} The following equivocation rates are
achievable for all $(\alpha,\beta_1,\beta_2)\in[0,1]^3$
\begin{align}
R_{e,1}&\leq \frac{1}{2}\log\left(1+\frac{\alpha
P(N_1+a_2\bar{\beta}_2P+N_2+N_{c,2})}{\bar{\alpha}P
(N_1+a_2\bar{\beta}_2P+N_2+N_{c,2})+(N_1+a_2\bar{\beta}_2P)(N_2+N_{c,2})}\right)\nonumber\\
&\quad -\frac{1}{2}\log\left(1+\alpha P
\left(\frac{1}{a_1\bar{\beta}_1P+N_2}+\frac{1}{N_1+N_{c,1}}\right)\right)\\
R_{e,2}&\leq \frac{1}{2}\log\left(1+\frac{\bar{\alpha}
P(N_2+a_1\bar{\beta}_1P+N_2+N_{c,1})}{\alpha P
(N_2+a_1\bar{\beta}_1P+N_1+N_{c,1})+(N_2+a_1\bar{\beta}_1P)(N_1+N_{c,1})}\right)\nonumber\\
&\quad -\frac{1}{2}\log\left(1+\alpha P
\left(\frac{1}{a_2\bar{\beta}_2P+N_1}+\frac{1}{N_2+N_{c,2}}\right)\right)
\end{align}
where
$\bar{\alpha}=1-\alpha,\bar{\beta}_1=1-\beta_1,\bar{\beta}_2=1-\beta_2$,
and $N_{c,1},N_{c,2}$ are subject to
\begin{align}
N_{c,1}&\geq
\frac{-b_{11}+\sqrt{b_{11}^2+4a_{11}c_{11}}}{2a_{11}}\\
N_{c,2}&\geq \frac{-b_{22}+\sqrt{b_{22}^2+4a_{22}c_{22}}}{2a_{22}}
\end{align}
and
\begin{align}
a_{11}&=a_1\beta_1 P \\
b_{11}&=P\big(P+a_1\beta_1(P+N_1)\big)-(P+N_1+a_2\bar{\beta}_2
P)(P+N_2+a_1\bar{\beta}_1 P)\\
c_{11}&=(P+N_1+a_2\bar{\beta}_2
P)\big(PN_1+(P+N_1)(N_2+a_1\bar{\beta}_1P)\big)\\
a_{22}&=a_2\beta_2 P \\
b_{22}&=P\big(P+a_2\beta_2(P+N_2)\big)-(P+N_1+a_2\bar{\beta}_2
P)(P+N_2+a_1\bar{\beta}_1 P)\\
c_{22}&=(P+N_2+a_1\bar{\beta}_1
P)\big(PN_2+(P+N_2)(N_1+a_2\bar{\beta}_2P)\big)
\end{align}
\end{Prop}

\begin{proof}
This achievable region is obtained by selecting $X=V_1+V_2$ where
$V_1\sim\mathcal{N}(0,\alpha P)$,
$V_2\sim\mathcal{N}(0,\bar{\alpha} P)$ and are independent,
$X_i=U_i+\tilde{Z}_i$ where $U_i\sim\mathcal{N}(0,a_i\beta_i P)$,
$\tilde{Z}_i\sim\mathcal{N}(0,a_i\bar{\beta}_i P),\break i=1,2$ and
independent, and $\hat{Y}_i=Y_i+Z_{c,i}$ where
$Z_{c,i}\sim\mathcal{N}(0,N_{c,i}), i=1,2$ and are independent of all
other random variables. Direct calculation of rates in
Theorem~\ref{Two_sided} with these random variable
selections yields the achievable region.
\end{proof}

\begin{figure}[t]
\begin{center}
\epsfxsize = 12.5 cm  \epsffile{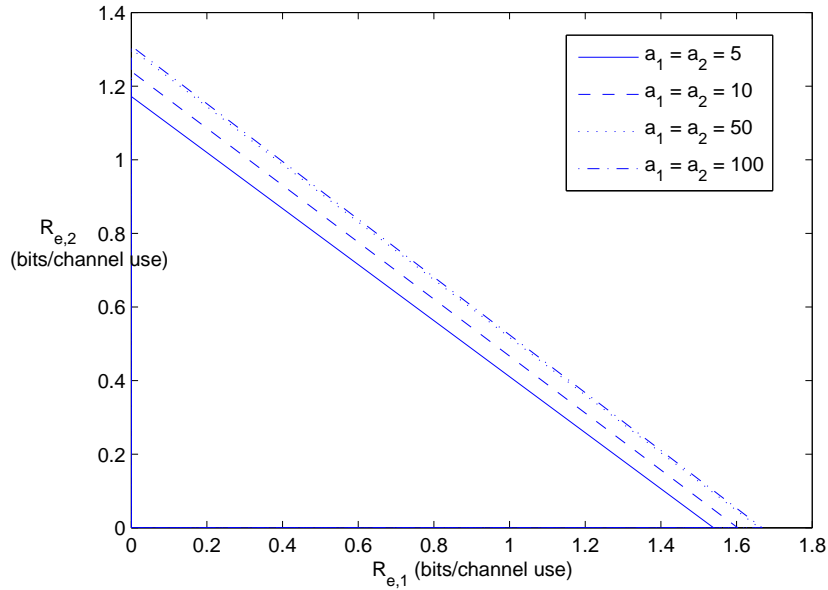}
\end{center}
\caption{Achievable equivocation rate region using
Proposition~\ref{Prop_TwoSided} where each user can jointly jam
and relay. $P=8, N_1=1, N_2=2$, i.e., user 2 cannot have any
positive secrecy in the underlying BC.} \label{Two_Sided_fig}
\end{figure}

A numerical example is given in Figure~\ref{Two_Sided_fig} for the
case $P=8,N_1=1,N_2=2$. Comparing Figure~\ref{Two_Sided_fig} with
Figures~
\ref{Jam_Relay_Ind_Input_User1_Stronger_fig} and~\ref{DPC_User1_Jam_Relay_User1_Stronger_fig},
we observe that user 2's secrecy rate improves significantly
because now user 2 can jam user 1 to improve its own secrecy rate. We
also observe that user 1's secrecy rate improves as well, compared
to Section~\ref{sec:Gaussian_Revisited}. The increase in user 1's secrecy in this
two-sided case is due to the fact that user 2 now acts as a relay for
user 1. However, when user 1 jams user 2 using all of its power,
it limits the help that comes from user 2, hence
Theorem~\ref{Two_sided} provides only a modest secrecy rate
increase for user 1 on top of what
Theorem~\ref{Jamming_and_Relaying_Thm} already provides.

\section{Conclusions}

In this paper, we investigated the effects of cooperation on
secrecy. We showed that user cooperation can increase secrecy,
i.e., even an untrusted party can help. An important point to
observe though is that whether cooperation can improve secrecy or
not depends on the cooperation method employed. For instance, even
though a decode-and-forward (DAF) based cooperation scheme can
increase the rate, it cannot improve secrecy, because in this case
the cooperating party, which is also the eavesdropper, needs to
decode the message it forwards. However, in CAF, we do not require
the cooperating party to decode the message. In fact, in CAF, the
cooperating party helps increase the rate of the main transmitter
to levels which it itself cannot decode, hence improving the
secrecy of the main transmitter-receiver pair against itself.

\appendices
\appendixpage

\section{Proof of Theorem \ref{Outer_Bound_Lemma}}
\label{Proof_of_outer_bound}

Here we prove the outer bound on the capacity-equivocation region
of the CRBC given in Theorem \ref{Outer_Bound_Lemma} which closely
follows the converse given in \cite{Korner} and the outer bound
in \cite{Ruoheng}. First, define the following random variables
\begin{eqnarray}
U_i &=& Y_1^{i-1}Y_{2,i+1}^n \label{Aux_U} \\
V_{1,i}&=& W_1 U_{i} \label{Aux_V1} \\
V_{2,i} &=& W_{2} U_i \label{Aux_V2}
\end{eqnarray}
which satisfy the following Markov chain
\begin{eqnarray}
U_i\rightarrow \left(V_{1,i},V_{2,i}\right) \rightarrow
\left(X_{i},X_{1,i},Y_{1,i}\right) \rightarrow Y_{2,i}
\end{eqnarray}
but do not satisfy the following one
\begin{eqnarray}
U_i\rightarrow \left(V_{1,i},V_{2,i}\right) \rightarrow
\left(X_{i},X_{1,i}\right) \rightarrow
\left(Y_{1,i},Y_{2,i}\right)
\end{eqnarray}
because of the encoding function employed at user 1 which can
generate correlation between $Y_{1,i} \textrm{ and }
\left(Y_{1,i+1}^n,Y_{2,i+1}^n\right)$ through $X_{1,i+1}$ that
cannot be resolved by conditioning on $(X_{i},X_{1,i})$. For a
similar discussion, the reader can refer to \cite{Kramer}.

We start with the achievable rate of user 1.
\begin{eqnarray}
nR_{1}&=& H(W_1)=I(W_1;Y_1^n)+H(W_1|Y_1^n)  \\
&\leq & I(W_1;Y_1^n)+\epsilon_n \label{Fanos_Lemma} \\
&=& \sum_{i=1}^n I(W_1;Y_{1,i}|Y_1^{i-1})+\epsilon_n  \\
&=& \sum_{i=1}^n H(W_1|Y_1^{i-1})-H(W_1|Y_1^{i-1},Y_{1,i})+\epsilon_n\\
&=& \sum_{i=1}^n H(W_1|Y_1^{i-1},X_{1,i})-H(W_1|Y_1^{i-1},Y_{1,i})+\epsilon_n \label{encoding_relay}\\
&\leq & \sum_{i=1}^n H(W_1|Y_1^{i-1},X_{1,i})-H(W_1|Y_1^{i-1},Y_{1,i},X_{1,i})+\epsilon_n \label{cond_cannot_inc3}\\
&=& \sum_{i=1}^n I(W_1;Y_{1,i}|Y_1^{i-1},X_{1,i})+\epsilon_n\\
&\leq& \sum_{i=1}^n H(Y_{1,i}|X_{1,i})-H(Y_{1,i}|Y_1^{i-1},X_{1,i},W_1)+\epsilon_n \label{cond_cannot_inc4}\\
&\leq& \sum_{i=1}^n H(Y_{1,i}|X_{1,i})-H(Y_{1,i}|Y_1^{i-1},X_{1,i},W_1,Y_{2,i+1}^n)+\epsilon_n \label{cond_cannot_inc5}\\
&=& \sum_{i=1}^n I(V_{1,i};Y_{1,i}|X_{1,i})+\epsilon_n \label{Definition_V1}
\end{eqnarray}
where (\ref{Fanos_Lemma}) is due to Fano's lemma,
(\ref{encoding_relay}) follows from the Markov chain
$W_1\rightarrow Y_1^{i-1}\rightarrow X_{1,i}$, (\ref{cond_cannot_inc3}),
(\ref{cond_cannot_inc4}) and (\ref{cond_cannot_inc5}) are due
to the fact that conditioning cannot increase entropy, and
(\ref{Definition_V1}) follows from the definition
of $V_{1,i}$ in (\ref{Aux_V1}). Similarly, for the achievable
rate of user 2, we have
\begin{eqnarray}
nR_2 &\leq & I(W_2;Y_2^n) +\epsilon_n \label{Fanos_Lemma1} \\
&=& \sum_{i=1}^n I(W_2;Y_{2,i}|Y_{2,i+1}^n)+\epsilon_n  \\
&=& \sum_{i=1}^n  H(Y_{2,i}|Y_{2,i+1}^n)-H(Y_{2,i}|Y_{2,i+1}^n,W_2)+\epsilon_n  \\
&\leq & \sum_{i=1}^n
H(Y_{2,i})-H(Y_{2,i}|Y_{2,i+1}^n,W_2,Y_1^{i-1})+\epsilon_n
\label{Conditioning_entropy1} \\
 &\leq & \sum_{i=1}^n
I(V_{2,i};Y_{2,i})+\epsilon_n \label{Definition_V2}
\end{eqnarray}
where (\ref{Fanos_Lemma1}) is due to Fano's lemma,
(\ref{Conditioning_entropy1}) is due to the fact that conditioning
cannot increase entropy, and (\ref{Definition_V2})
follows from the definition of $V_{2,i}$ given in (\ref{Aux_V2}).

We now derive the outer bounds on the equivocation rates. We start
with user 1.
\begin{eqnarray}
nR_{e,1} &=&H(W_1|Y_2^n)=H(W_1)-I(W_1;Y_2^n)  \\
&=& I(W_1;Y_1^n)-I(W_1;Y_2^n)+H(W_1|Y_1^n)  \\
& \leq & I(W_1;Y_1^n)-I(W_1;Y_2^n)+\epsilon_n \label{Fanos_Lemma2} \\
&=& \sum_{i=1}^n I(W_1;Y_{1,i}|Y_1^{i-1})-
I(W_1;Y_{2,i}|Y_{2,i+1}^n) +\epsilon_n  \\
&=& \sum_{i=1}^n
I(W_1,Y_{2,i+1}^n;Y_{1,i}|Y_1^{i-1})-I(Y_{2,i+1}^n;Y_{1,i}|Y_1^{i-1},W_1)
-I(W_1,Y_{1}^{i-1};Y_{2,i}|Y_{2,i+1}^n)\nonumber \\
&&\quad + I(Y_{1}^{i-1};Y_{2,i}|Y_{2,i+1}^n,W_1) +\epsilon_n
\label{dummy_step}
\end{eqnarray}
where (\ref{Fanos_Lemma2}) is due to Fano's lemma. Using~\cite{Korner}
\begin{equation}
\sum_{i=1}^n I(Y_{2,i+1}^n;Y_{1,i}|Y_1^{i-1},W_1) =\sum_{i=1}^n
I(Y_{1}^{i-1};Y_{2,i}|Y_{2,i+1}^n,W_1)
\end{equation}
in (\ref{dummy_step}), we obtain
\begin{eqnarray}
nR_{e,1} &\leq & \sum_{i=1}^n I(W_1,Y_{2,i+1}^n;Y_{1,i}|Y_1^{i-1})
-I(W_1,Y_{1}^{i-1};Y_{2,i}|Y_{2,i+1}^n)+\epsilon_n  \\
&=& \sum_{i=1}^n I(W_1;Y_{1,i}|Y_1^{i-1},Y_{2,i+1}^n)+
I(Y_{2,i+1}^n;Y_{1,i}|Y_1^{i-1})
-I(W_1;Y_{2,i}|Y_{2,i+1}^n,,Y_{1}^{i-1})\nonumber\\
&&\quad -I(Y_{1}^{i-1};Y_{2,i}|Y_{2,i+1}^n)+\epsilon_n
\label{dummy_step_xx}
\end{eqnarray}
Now, using \cite{Korner}
\begin{equation}
\sum_{i=1}^n I(Y_{2,i+1}^n;Y_{1,i}|Y_1^{i-1}) =\sum_{i=1}^n
I(Y_{1}^{i-1};Y_{2,i}|Y_{2,i+1}^n)
\end{equation}
in (\ref{dummy_step_xx}), we obtain
\begin{eqnarray}
nR_{e,1} &\leq & \sum_{i=1}^n I(W_1;Y_{1,i}|Y_1^{i-1},Y_{2,i+1}^n)
-I(W_1;Y_{2,i}|Y_{2,i+1}^n,Y_{1}^{i-1})+\epsilon_n\\
&=& \sum_{i=1}^n I(W_{1};Y_{1,i}|U_{i})
-I(W_{1};Y_{2,i}|U_{i})+\epsilon_n \label{definition_of_u}\\
&=& \sum_{i=1}^n I(W_{1},U_{i};Y_{1,i}|U_{i})
-I(W_{1},U_{i};Y_{2,i}|U_{i})+\epsilon_n\\
&=& \sum_{i=1}^n I(V_{1,i};Y_{1,i}|U_{i})
-I(V_{1,i};Y_{2,i}|U_{i})+\epsilon_n
\label{definition_of_v1}
\end{eqnarray}
where (\ref{definition_of_u}) and
(\ref{definition_of_v1}) follow from the definitions of $U_i$ and
$V_{1,i}$ given in (\ref{Aux_U}) and (\ref{Aux_V1}), respectively.
Similarly, we can use the preceding technique for user 2's equivocation
rate as well after noting that
\begin{eqnarray}
nR_{e,2} &\leq & H(W_2|Y_1^n,X_1^n)\leq H(W_2|Y_1^n)
\end{eqnarray}
which leads to
\begin{eqnarray}
nR_{e,2} &\leq &  \sum_{i=1}^n I(V_{2,i};Y_{2,i}|U_{i})
-I(V_{2,i};Y_{1,i}|U_{i})+\epsilon_n
\end{eqnarray}

The other bounds on the equivocation rates can be derived as
follows.
\begin{align}
nR_{e,1}&= H(W_1|Y_2^n)\leq H(W_1,W_2|Y_2^n) \\
&= H(W_1|W_2,Y_2^n)+H(W_2|Y_2^n) \\
&\leq  H(W_1|W_2,Y_2^n)+\epsilon_n \label{Fanos_Lemma3}  \\
&= I(W_1;Y_1^n|W_2)-I(W_1;Y_2^n|W_2)+H(W_1|W_2,Y_1^n)+\epsilon_n
 \\
&\leq  I(W_1;Y_1^n|W_2)-I(W_1;Y_2^n|W_2)+\epsilon_n^{\prime}
\label{Fanos_Lemma4}
 \\
&= \sum_{i=1}^n
I(W_1;Y_{1,i}|W_2,Y_{1}^{i-1})-I(W_1;Y_{2,i}|W_2,Y_{2,i+1}^n)+\epsilon_n^{\prime}
 \\
&= \sum_{i=1}^n
I(W_1,Y_{2,i+1}^n;Y_{1,i}|W_2,Y_{1}^{i-1})-I(W_1,Y_{1}^{i-1};Y_{2,i}|W_2,Y_{2,i+1}^n)+\epsilon_n^{\prime}
\label{Csiszar_identity_1}
 \\
&= \sum_{i=1}^n
I(W_1;Y_{1,i}|W_2,Y_{1}^{i-1},Y_{2,i+1}^n)-I(W_1;Y_{2,i}|W_2,Y_{2,i+1}^n,Y_{1}^{i-1})+\epsilon_n^{\prime}
\label{Csiszar_identity_2}
 \\
 &= \sum_{i=1}^n
I(W_1;Y_{1,i}|W_2,U_{i})-I(W_1;Y_{2,i}|W_2,U_{i})+\epsilon_n^{\prime}
\label{definition_of_u_1}
 \\
  &= \sum_{i=1}^n
I(W_1,U_i;Y_{1,i}|W_2,U_{i})-I(W_1,U_i;Y_{2,i}|W_2,U_{i})+\epsilon_n^{\prime}
 \\
&= \sum_{i=1}^n
I(V_{1,i};Y_{1,i}|V_{2,i})-I(V_{1,i};Y_{2,i}|V_{2,i})+\epsilon_n^{\prime}
\label{definition_of_v1_v2}
\end{align}
where (\ref{Fanos_Lemma3}) and (\ref{Fanos_Lemma4}) are due to
Fano's lemma, and (\ref{Csiszar_identity_1}) and
(\ref{Csiszar_identity_2}) are due to the following identities~\cite{Korner}
\begin{align}
\sum_{i=1}^nI(Y_{2,i+1}^n;Y_{1,i}|W_1,W_2,Y_{1}^{i-1})&=\sum_{i=1}^nI(Y_{1}^{i-1};Y_{2,i}|W_1,W_2,Y_{2,i+1}^n)\\
\sum_{i=1}^nI(Y_{2,i+1}^n;Y_{1,i}|W_2,Y_{1}^{i-1})&=\sum_{i=1}^nI(Y_{1}^{i-1};Y_{2,i}|W_2,Y_{2,i+1}^n)
\end{align}
respectively. Finally, (\ref{definition_of_u_1}) and
(\ref{definition_of_v1_v2}) follow from the definitions of
$U_i,V_{1,i}\textrm{ and }V_{2,i}$ given in (\ref{Aux_U}),
(\ref{Aux_V1}) and (\ref{Aux_V2}), respectively. Similarly, we
can use this technique to bound user 2's equivocation
rate after noting that $H(W_2|Y_1^n,X_1^n)\leq H(W_2|Y_1^n)$, which
leads to
\begin{align}
nR_{e,2} \leq  H(W_2|Y_1^n,X_1^n)\leq H(W_2|Y_1^n)\leq \sum_{i=1}^n
I(V_{2,i};Y_{2,i}|V_{1,i})-I(V_{2,i};Y_{2,i}|V_{1,i})+\epsilon_n^{\prime}
\end{align}

To express the outer bounds obtained above in a single-letter
form, we define
$U=JU_{J},V_1=V_{1,J},V_2=V_{2,J},X=X_{J},X_1=X_{1,J},Y_1=Y_{1,J},Y_2=Y_{2,J}$
where $J$ is a random variable which is uniformly distributed over
$\{1,\ldots,n\}$. Using these new definitions, we can reach the
single-letter expressions given in
Theorem~\ref{Outer_Bound_Lemma}, hence completing the proof.

\section{Proof of Theorem~\ref{cor_Re2}}
\label{proof_simpler_outer_bound} The proof is as follows.
\begin{align}
R_{e,2}\leq H(W_2|Y_1^n,X_1^n)&\leq I(W_2;Y_2^n|X_1^n)-I(W_2;Y_1^n|X_1^n)+H(W_2|Y_2^n,X_1^n)\\
&\leq I(W_2;Y_2^n|X_1^n)-I(W_2;Y_1^n|X_1^n)+\epsilon_n \label{Fanos_Lemma5}\\
&\leq I(W_2;Y_2^n|X_1^n,Y_1^n)+\epsilon_n\\
&\leq I(X^n,W_2;Y_2^n|X_1^n,Y_1^n)+\epsilon_n \\
&= I(X^n;Y_2^n|X_1^n,Y_1^n)+\epsilon_n \label{encoding}\\
&=\sum_{i=1}^n I(X^n;Y_{2,i}|X_1^n,Y_1^n,Y_2^{i-1})+\epsilon_n\\
&\leq \sum_{i=1}^n H(Y_{2,i}|X_{1,i},Y_{1,i})-H(Y_{2,i}|X_1^n,Y_1^n,Y_2^{i-1},X^n)+\epsilon_n \label{cond_cannot_inc2}\\
&= \sum_{i=1}^n H(Y_{2,i}|X_{1,i},Y_{1,i})-H(Y_{2,i}|X_{1,i},Y_{1,i},X_{i})+\epsilon_n \label{memoryless_etc} \\
&= \sum_{i=1}^n I(X_i;Y_{2,i}|X_{1,i},Y_{1,i})+\epsilon_n
\end{align}
where (\ref{Fanos_Lemma5}) is due to Fano's lemma,
(\ref{encoding}) follows from the fact that given $X^n$, $W_2$ is
independent of all other random variables,
(\ref{cond_cannot_inc2}) is due to the fact that conditioning
cannot increase entropy, and (\ref{memoryless_etc}) follows
from the Markov chains
\begin{align}
(Y_{1,i},Y_{2,i})\rightarrow (X_i,X_{1,i})\rightarrow (Y_1^{i-1},Y_2^{i-1},X^{i-1},X_1^{i-1})\\
Y_{2,i}\rightarrow (X_i,X_{1,i},Y_{1,i})\rightarrow
(Y_{1,i+1}^n,X_{i+1}^n,X_{1,i+1}^n)
\end{align}
Thus, after defining an independent random variable $J$, that is
uniformly distributed over $\{1,\ldots,n\}$, and
$X=X_{J},X_1=X_{1,J},Y_1=Y_{1,J},Y_{2}=Y_{2,J}$, we can obtain the
single-letter expression in Theorem~\ref{cor_Re2}, completing the proof.

\section{Proof of Corollary~\ref{Cor_a_infinity}}
\label{Proof_a_infinity}

In Propositions~\ref{Independent_Inputs} and
\ref{Dirty_paper_User1}, if we take $a\rightarrow\infty$, then the
secrecy rate in (\ref{max_secrecy_a_infinity}) can be shown to be
achievable. As a notational remark, $H(\cdot)$ denotes the
differential entropy in this section. We now compute an outer
bound for $R_{e,2}$ using Theorem~\ref{cor_Re2},
\begin{align}
R_{e,2}&\leq I(X;Y_2|X_1,Y_1)\\
&=H(Y_2|X_1,Y_1)-H(Z_2|Z_1)\\
&\leq H(X+Z_2|Y_1)-H(Z_2)\label{cond_cannot_inc}\\
&\leq H(X+Z_2-\alpha Y_1)-\frac{1}{2}\log(2\pi eN_2) \label{cond_cannot_inc1}\\
&\leq \frac{1}{2}\log (2\pi e) E\left[(X+Z_2-\alpha
Y_1)^2\right]-\frac{1}{2}\log(2\pi eN_2)\label{gauss_maximize}\\
&\leq \frac{1}{2}\log \left((1-\alpha)^2P+\alpha^2
N_1+N_2\right)-\frac{1}{2}\log(N_2)\label{use_power_const}
\end{align}
where in (\ref{cond_cannot_inc}), we used the fact that
conditioning cannot increase entropy and that $H(Z_2|Z_1)\break
=H(Z_2)$ due to the independence of $Z_1$ and $Z_2$. Equation
(\ref{cond_cannot_inc1}) is again due to the fact that
conditioning cannot increase entropy, (\ref{gauss_maximize}) comes
from the fact that Gaussian distribution maximizes entropy
subject to a power constraint, and (\ref{use_power_const}) is
obtained by using the power constraint on $X$. Finally, we note
that (\ref{use_power_const}) is a valid outer bound for every
$\alpha$ and if we select $\alpha$ as
\begin{align}
\alpha=\frac{P}{P+N_1}
\end{align}
we get (\ref{max_secrecy_a_infinity}), completing the proof.

\section{Proof of Theorem \ref{Jamming_and_Relaying_Thm}}
\label{Proof_of_Jamming_and_Relaying_Thm}

The transmitter uses the joint encoding scheme of Marton
\cite{Marton} and user 1 uses a CAF scheme \cite{Cover}. User 2
employs list decoding to find which $\hat{Y}_1$ is sent. Let
$A_{\epsilon}^{n}(V_1)\textrm{ and }A_{\epsilon}^{n}(V_2)$ denote
the sets of strongly typical i.i.d. length-$n$ sequences of
$\bold{v}_1 \textrm{ and }\bold{v}_2$, respectively. Let
$A_{\epsilon}^{n}\left(V_1|\bold{v}_2\right)\left(\textrm{resp.
}A_{\epsilon}^{n}\left(V_2|\bold{v}_1\right)\right)$ denote the
set of length-$n$ sequences $V_1$ (resp. $V_2$) that are jointly
typical with $\bold{v}_2\textrm{ (resp. }\bold{v}_1)$.
Furthermore, let $S_{\epsilon}^{n}(\bold{v}_1)\textrm{ (resp.
}S_{\epsilon}^{n}(\bold{v}_2))$ denote the set of
$\bold{v}_1\textrm{ (resp. }\bold{v}_2)$ sequences for which
$A_{\epsilon}^{n}\left(V_2|\bold{v}_1\right) \left(\textrm{resp.
}A_{\epsilon}^{n}\left(V_1|\bold{v}_2\right)\right)$ are
non-empty. Fix the probability distribution as
\begin{equation}
p(v_1,v_2)p(x|v_1,v_2)p(u,x_1)p(\hat{y}_1|u,v_1,y_1)
\end{equation}

\noindent \underline{\textbf{Codebook structure:}}
\begin{enumerate}
    \item Select $2^{nR(V_i)}$ $\bold{v}_i$ sequences
    through
    \begin{equation}
    p(\bold{v}_i)=\left\{\begin{array}{ll} \frac{1}{||S_{\epsilon}^{n}(\bold{v}_i)||},   & \textrm{if } \bold{v}_i\in S_{\epsilon}^{n}(\bold{v}_i)
    \vspace{0.2cm} \\
    0,   & \textrm{otherwise }
    \end{array} \right.
    \end{equation}
    in an i.i.d. manner and index them as $\bold{v}_i (w_i,\tilde{w}_i,l_i)$ where
    $w_i\in\left\{1,\ldots,2^{nR_i}\right\}$,
    $\tilde{w}_i\in\{1,\ldots,2^{n\tilde{R}_i}\}$ and
    $l_i\in\left\{1,\ldots,2^{nL_i}\right\}$ for $i=1,2$.
    $R_i,\tilde{R}_i,L_i$ and $R(V_i)$ are related through
    \begin{align}
    R(V_i)=R_i+\tilde{R}_i+L_i,\qquad i=1,2
    \end{align}
    Furthermore, we set
    \begin{align}
    L_1+L_2=I(V_1;V_2)+\epsilon
    \label{ensure_typical_pair}
    \end{align}
    to ensure that for given pairs $(w_1,\tilde{w}_1)$ and
    $(w_2,\tilde{w}_2)$, we can find a jointly typical pair
    $(\bold{v}_1(w_1,\tilde{w}_1,l_1),\bold{v}_2(w_2,\tilde{w}_2,l_2))$
    for some $l_1,l_2$.

    \item For each $(w_1,w_2)$, the transmitter randomly picks
    $(\tilde{w}_1,\tilde{w}_2)$ and finds a pair
    \begin{math}(\bold{v}_1(w_1,\tilde{w}_1,l_1),\newline\bold{v}_2(w_2,\tilde{w}_2,l_2))\end{math}
    that is jointly typical. Such a pair exists with high
    probability due to (\ref{ensure_typical_pair}). Then, given
    this pair of $(\bold{v}_1,\bold{v}_2)$, the transmitter generates
    its channel inputs through $\prod_{i=1}^n p(x_i
    |v_{1,i},v_{2,i})$.

    \item User 1 generates $2^{nR_0}$ length-$n$ sequences $\bold{u}$ through
    $p(\bold{u})=\prod_{i=1}^{n}p(u_i)$
    and labels them as $\bold{u}(s_i)$ where $s_i\in
    \{1,\ldots,2^{nR_0}\}$.

    \item For each $\bold{u}(s_i)$, user 1 generates
    $2^{n\hat{R}}$ length-$n$ sequences $\hat{\bold{y}}_1$ through
    $p(\hat{\bold{y}}_1|\bold{u})=\prod_{i=1}^n \break
    p(\hat{y}_{1,i}|u_i)$ and indexes them as $\hat{\bold{y}}_1
    (z_i|s_i)$ where $z_i\in\{1,\ldots,2^{n\hat{R}}\}$.

    \item For each $\bold{u}(s_i)$, user 1 generates
    $2^{nR_0^{\prime}}$ length-$n$ sequences $\bold{x}_1$ through
    $p(\bold{x}_1|\bold{u})=\prod_{i=1}^n\break p(x_{1,i}|u_i)$ and
    indexes them as $\bold{x}_1(t_i|s_i)$ where
    $t_i\in\{1,\ldots,2^{nR_0^{\prime}}\}$.

\end{enumerate}

\noindent\underline{\textbf{Partitioning:}}

\begin{itemize}
    \item Partition $2^{n\hat{R}}$ into cells $S_{s_i}$ where
$s_i\in\{1,\ldots,2^{nR_0}\}$.
\end{itemize}

\noindent\underline{\textbf{Encoding:}}

\vspace{0.2cm} The transmitter sends $\bold{x}$ corresponding to
the pair $(w_1,w_2)$. User 1 (relay) sends $\bold{x}_1(t_i|s_i)$
if the estimate of $\bold{y}_{1}(i-1)$, i.e., $\hat{z}_{i-1}$,
falls into $S_{s_i}$ and $t_i$ is chosen randomly from
$\{1,\ldots,2^{nR_0^{\prime}}\}$. The use of many
$\bold{x}_1(t_i|s_i)$ for actual help signal $\bold{u}(s_i)$ aims
to confuse user 2 and to decrease its decoding capability.

\vspace{0.5cm}

\noindent\underline{\textbf{Decoding:}}

\vspace{0.25cm} \noindent\underline{\textbf{a. Decoding at user
1:}}
\begin{enumerate}
    \item User 1 seeks a unique typical pair of
    $\left(\bold{y}_1(i),\bold{v}_{1}(w_{1,i},\tilde{w}_{1,i},l_i),\bold{x}_1(t_i|s_i)\right)$
    which can be achieved with vanishingly small error probability if
    \begin{equation}
    \label{rate_1}
    R(V_1)\leq I(V_1;Y_1|X_1)
    \end{equation}
    \item User 1 decides that $z_i$ is received if there exists a
    jointly typical pair $(\hat{\bold{y}}_{1}(z_i|s_i),\bold{y}_{1}(i),\break\bold{v}_{1}(w_{1,i},\tilde{w}_{1,i},l_i),\bold{x}_1(t_i|s_i))$
    which can be guaranteed to occur if
    \begin{equation}
    \label{Compression_constraint_1}
    \hat{R}\geq I(\hat{Y}_1;Y_1 |U,X_1,V_1)
    \end{equation}
\end{enumerate}

\noindent\underline{\textbf{b. Decoding at user 2:}}

\begin{enumerate}
    \item User 2 seeks a unique jointly typical pair of
    $\left(\bold{y}_2(i),\bold{u}(s_i)\right)$
    which can be found with vanishingly small error probability if
    \begin{equation}
    R_0\leq I(U;Y_2)
    \end{equation}
    \item User 2 employs list decoding to decode
    $\bold{\hat{y}}_1(z_{i-1}|s_{i-1})$. It first calculates its
    ambiguity set as
    \begin{equation}
\mathcal{L}\left(\hat{\bold{y}}_1(z_{i-1}|\hat{s}_{i-1})\right)=\left\{\hat{\bold{y}}_1(z_{i-1}|\hat{s}_{i-1}):\left(\hat{\bold{y}}_1(z_{i-1}|\hat{s}_{i-1}),\bold{y}_{2}(i-1)\right)\textrm{ is jointly typical}  \right\}
    \end{equation}
    and takes its intersection with $S_{\hat{s}_i}$ which results
    in a unique and correct intersection point if
    \begin{equation}
    \label{Compression_constraint_2}
    \hat{R}\leq
    I(\hat{Y}_1;Y_2|U)+R_0\leq I(\hat{Y}_1,U;Y_2)
    \end{equation}
    Equations (\ref{Compression_constraint_1}) and
    (\ref{Compression_constraint_2}) lead to the compression
    constraint in (\ref{compression_constraint_jam_relay}).
    \item User 2 decides that $\bold{v}_2(w_{2,i-1},\tilde{w}_{2,i-1},l_{2,i-1})$ is received if
    there exists a unique jointly typical pair
    $\left(\bold{v}_2(w_{2,i-1},\tilde{w}_{2,i-1},l_{2,i-1}),\bold{y}_{2}(i-1),\hat{\bold{y}}_{1}(\hat{z}_{i-1}|\hat{s}_{i-1})\right)$,
    which can be found with vanishingly small error probability
    if
    \begin{equation}
    \label{rate_2}
    R(V_2)\leq I(V_2;Y_2,\hat{Y}_1|U)
    \end{equation}
\end{enumerate}

\noindent\underline{\textbf{Equivocation computation:}}

\vspace{0.5cm}

We now show that $R_{e,1}$ and $R_{e,2}$ satisfying
(\ref{equi_user_1_jamming_relaying_2})-(\ref{equi_user_1_jamming_relaying_1})
and
(\ref{equi_user_2_jamming_relaying_2})-(\ref{equi_user_2_jamming_relaying_1})
are achievable with the coding scheme presented. To this end, we
treat several possible cases separately. First, assume that
\begin{align}
R_1&\geq I(V_1;Y_1|X_1)-I(V_1;Y_2,\hat{Y}_1|V_2,U)-I(V_1;V_2)\label{case1_1}\\
R_2&\geq
I(V_2;Y_2,\hat{Y}_1|U)-I(V_2;Y_1|V_1,X_1)-I(V_1;V_2)\label{case1_2}
\end{align}
For this case, we select the total number of codewords, i.e.,
$R(V_i), i=1,2$, as
\begin{align}
R(V_1)&=I(V_1;Y_1|X_1)\label{total_number_codewords_1}\\
R(V_2)&=I(V_2;Y_2,\hat{Y}_1|U)\label{total_number_codewords_2}
\end{align}
With this selection, we have
\begin{align}
\tilde{R}_1+L_1&\leq I(V_1;Y_2,\hat{Y}_1|V_2,U)+I(V_1;V_2)\label{redundancy_1}\\
\tilde{R}_2+L_2 & \leq I(V_2;Y_1|V_1,X_1)+I(V_1;V_2)
\label{redundancy_2}
\end{align}
We start with user 1's equivocation rate,
\begin{align}
H(W_1|Y_2^n)&\geq H(W_1|Y_2^n,V_2^n,U^n,\hat{Y}_1^n) \label{equi_user1_start}\\
&=
H(W_1,Y_2^n,V_2^n,\hat{Y}_1^n|U^n)-H(Y_2^n,V_2^n,\hat{Y}_1^n|U^n)\\
&
=H(V_1^n,W_1,Y_2^n,V_2^n,\hat{Y}_1^n|U^n)-H(V_1^n|W_1,Y_2^n,V_2^n,\hat{Y}_1^n,U^n)\nonumber\\
&\quad -H(Y_2^n,V_2^n,\hat{Y}_1^n|U^n) \\
&=H(V_1^n|U^n)+H(W_1,Y_2^n,V_2^n,\hat{Y}_1^n|U^n,V_1^n)-H(V_1^n|W_1,Y_2^n,V_2^n,\hat{Y}_1^n,U^n)\nonumber\\
&\quad -H(Y_2^n,V_2^n,\hat{Y}_1^n|U^n) \\
&\geq
H(V_1^n|U^n)-I(V_1^n;Y_2^n,V_2^n,\hat{Y}_1^n|U^n)-H(V_1^n|W_1,Y_2^n,V_2^n,\hat{Y}_1^n,U^n)
\label{equi_user1_step1}
\end{align}
where each term will be treated separately. First term is
\begin{align}
H(V_1^n|U^n)=H(V_1^n)=nR(V_1)=nI(V_1;Y_1|X_1)
\label{equi_user1_step2}
\end{align}
where the first equality is due to the independence of $U^n$ and
$V_1^n$. The second equality follows from the fact that $V_1^n$
can take $2^{nR(V_1)}$ values with equal probability. The third
equality comes from our selection in
(\ref{total_number_codewords_1}). The second term of
(\ref{equi_user1_step1}) can be bounded as
\begin{align}
I(V_1^n;Y_2^n,V_2^n,\hat{Y}_1^n|U^n)\leq n
I(V_1;Y_2,V_2,\hat{Y}_1|U)+n\epsilon_n \label{equi_user1_step3}
\end{align}
using the approach devised in Lemma~3 of \cite{Ruoheng}. To bound
the last term in (\ref{equi_user1_step1}), we assume that user 2
is trying to decode $V_1^n$ given the side information $W_1=w_1$.
Since $V_1^n$ can take less than
$2^{n(I(V_1;Y_2,\hat{Y}_1|U,V_2)+I(V_1;V_2))}$ values (see
(\ref{redundancy_1})) given $W_1=w_1$, user 2 can decode $V_1^n$
with vanishingly small error probability as long as $W_1=w_1$ is
given. Consequently, the use of Fano's lemma yields
\begin{align}
H(V_1^n|W_1,Y_2^n,V_2^n,\hat{Y}_1^n,U^n)\leq \epsilon_n
\label{equi_user1_step4}
\end{align}
Plugging (\ref{equi_user1_step2}), (\ref{equi_user1_step3}) and
(\ref{equi_user1_step4}) into (\ref{equi_user1_step1}), we get
\begin{align}
H(W_1|Y_2^n)&\geq
nI(V_1;Y_1|X_1)-nI(V_1;Y_2,\hat{Y}_1,V_2|U)-n\epsilon_n\\
&=nI(V_1;Y_1|X_1)-nI(V_1;Y_2,\hat{Y}_1|V_2,U)-nI(V_1;V_2)-n\epsilon_n
\label{independence_u_v1v2}
\end{align}
where (\ref{independence_u_v1v2}) follows from the independence
of $(V_1,V_2)$ and $U$, i.e., $I(V_1;V_2|U)=I(V_1;V_2)$.
Similarly, we can bound equivocation of user 2 as follows,
\begin{align}
H(W_2|Y_1^n,X_1^n)&\geq H(W_2|Y_1^n,X_1^n,V_1^n) \\
&=H(W_2,Y_1^n,V_1^n|X_1^n)-H(Y_1^n,V_1^n|X_1^n)\\
&=H(W_2,V_2^n,Y_1^n,V_1^n|X_1^n)-H(V_2^n|W_2,Y_1^n,V_1^n,X_1^n)-H(Y_1^n,V_1^n|X_1^n)\\
&=H(V_2^n|X_1^n)+H(W_2,Y_1^n,V_1^n|X_1^n,V_2^n)-H(V_2^n|W_2,Y_1^n,V_1^n,X_1^n)\nonumber\\
&\quad -H(Y_1^n,V_1^n|X_1^n)\\
&\geq
H(V_2^n|X_1^n)-I(V_2^n;Y_1^n,V_1^n|X_1^n)-H(V_2^n|W_2,Y_1^n,V_1^n,X_1^n)
\label{equi_user2_step1}
\end{align}
where the first term is
\begin{align}
H(V_2^n|X_1^n)=H(V_2^n)=nR(V_2)=nI(V_2;Y_2,\hat{Y}_1|U)
\label{equi_user2_step2}
\end{align}
where the first equality is due to the independence of $V_2^n$ and
$X_1^n$, the second equality comes from the fact that $V_2^n$ can
take $2^{nR(V_2)}$ values with equal probability and the last
equality is a consequence of our choice in
(\ref{total_number_codewords_2}). The second term of
(\ref{equi_user2_step1}) can be bounded as
\begin{align}
I(V_2^n;Y_1^n,V_1^n|X_1^n)\leq nI(V_2;Y_1,V_1|X_1)+n\epsilon_n
\label{equi_user2_step3}
\end{align}
following the approach of Lemma~3 of \cite{Ruoheng}. To bound the
last term of (\ref{equi_user2_step1}), we assume that user 1 is
trying to decode $V_2^n$ given the side information $W_2=w_2$.
Since $V_2^n$ can take at most
$2^{n(I(V_2;Y_1|V_1,X_1)+I(V_2;V_1))}$ values (see
(\ref{redundancy_2})) given $W_2=w_2$, user 1 can decode $V_2^n$
with vanishingly small error probability as long as this side
information is available. Consequently, the use of Fano's lemma
yields
\begin{align}
H(V_2^n|W_2,Y_1^n,V_1^n,X_1^n)\leq \epsilon_n
\label{equi_user2_step4}
\end{align}
Plugging (\ref{equi_user2_step2}), (\ref{equi_user2_step3}) and
(\ref{equi_user2_step4}) into (\ref{equi_user2_step1}), we get
\begin{align}
H(W_2|Y_1^n,X_1^n)&\geq
nI(V_2;Y_2,\hat{Y}_1|U)-nI(V_2;Y_1,V_1|X_1)-n\epsilon_n \\
&= nI(V_2;Y_2,\hat{Y}_1|U)-nI(V_2;Y_1|V_1,X_1)-nI(V_1;V_2)-n\epsilon_n
\label{independence_x1_v1v2}
\end{align}
where (\ref{independence_x1_v1v2}) follows from the independence
of $(V_1,V_2)$ and $X_1$, i.e., $I(V_1;V_2|X_1)=I(V_1;V_2)$.

We have completed the equivocation calculation for the case
described by (\ref{case1_1})-(\ref{case1_2}). The proofs of other
cases involve no different arguments besides decreasing the total
number codewords in
(\ref{total_number_codewords_1})-(\ref{total_number_codewords_2}).
For example, if
\begin{align}
R_1\leq I(V_1;Y_1|X_1)-I(V_1;Y_2,\hat{Y}_1|V_2,U)-I(V_1;V_2)
\end{align}
then we select the total number of codewords for user 1 as
\begin{align}
R(V_1)=R_1+I(V_1;Y_2,\hat{Y}_1|V_2,U)+I(V_1;V_2)\label{total_number_codewords_1_case1}
\end{align}
which is equivalent to saying that
\begin{align}
\tilde{R}_1+L_1=I(V_1;Y_2,\hat{Y}_1|V_2,U)+I(V_1;V_2)
\end{align}
In this case, following the steps from (\ref{equi_user1_start}) to
(\ref{equi_user1_step1}), we can bound the equivocation of user 1
as follows,
\begin{align}
H(W_1|Y_2^n)\geq
H(V_1^n|U^n)-I(V_1^n;Y_2^n,V_2^n,\hat{Y}_1^n|U^n)-H(V_1^n|W_1,Y_2^n,V_2^n,\hat{Y}_1^n,U^n)
\label{equi_user1_step1_case2}
\end{align}
where the first term is now
\begin{align}
H(V_1^n|U^n)=H(V_1^n)=nR(V_1)=n(R_1+I(V_1;Y_2,\hat{Y}_1|V_2,U)+I(V_1;V_2))
\label{equi_user1_step2_case2}
\end{align}
where the first equality is due to the independence of $V_1^n$ and
$U^n$, the second equality is due to the fact that $V_1^n$ can
take at most $2^{nR(V_1)}$ values with equal probability and the last
equality is a consequence of our choice in
(\ref{total_number_codewords_1_case1}). An upper bound on the
second term was already obtained in (\ref{equi_user1_step3}). The
third term can also be shown to decay to zero as $n$ goes to
infinity considering the case that user 2 is decoding $V_1^n$
using side information $W_1=w_1$. Since $V_1^n$ can take
$2^{n(I(V_1;Y_2,\hat{Y}_1|V_2,U)+I(V_1;V_2))}$ values given
$W_1=w_1$, user 2 can decode $V_2^n$ with vanishingly small error
probability as long as this side information is available.
Therefore, the use of Fano's lemma implies
\begin{align}
H(V_1^n|W_1,Y_2^n,V_2^n,\hat{Y}_1^n,U^n)\leq \epsilon_n
\label{equi_user1_step3_case2}
\end{align}
Plugging (\ref{equi_user1_step3}),(\ref{equi_user1_step2_case2}),
(\ref{equi_user1_step3_case2}) into
(\ref{equi_user1_step1_case2}), we get
\begin{align}
H(W_1|Y_2^n)&\geq
n(R_1+I(V_1;Y_2,\hat{Y}_1|V_2,U)+I(V_1;V_2))-I(V_1;Y_2,V_2,\hat{Y}_1|U)-n\epsilon_n
\\
&= nR_1-n\epsilon_n
\end{align}
where we used the fact that $U$ and $(V_1,V_2)$ are independent,
i.e., $I(V_1;V_2|U)=I(V_1;V_2)$. The other cases leading to
different equivocation rates can be proved similarly, hence
omitted.

\section{Proof of Theorem~\ref{Two_sided}}
\label{Proof_of_two_sided}

Fix the probability distribution as
\begin{equation}
p(v_1,v_2)p(x|v_1,v_2)p(u_1,x_1)p(\hat{y}_1|u_1,y_1)p(u_2,x_2)p(\hat{y}_2|u_2,y_2)
\end{equation}

\noindent \underline{\textbf{Codebook structure:}}
\begin{enumerate}
    \item Select $2^{nR(V_i)}$ $\bold{v}_i$ sequences
    through
    \begin{equation}
    p(\bold{v}_i)=\left\{\begin{array}{ll} \frac{1}{||S_{\epsilon}^{n}(\bold{v}_i)||},   & \textrm{if } \bold{v}_i\in S_{\epsilon}^{n}(\bold{v}_i)
    \vspace{0.2cm} \\
    0,   & \textrm{otherwise }
    \end{array} \right.
    \end{equation}
    in an i.i.d. manner and index them as $\bold{v}_i (w_i,\tilde{w}_i,l_i)$ where
    $w_i\in\left\{1,\ldots,2^{nR_i}\right\}$,
    $\tilde{w}_i\in\{1,\ldots,2^{n\tilde{R}_i}\}$ and
    $l_i\in\left\{1,\ldots,2^{nL_i}\right\}$ for $i=1,2$.
    $R_i,\tilde{R}_i,L_i$ and $R(V_i)$ are related through
    \begin{align}
    R(V_i)=R_i+\tilde{R}_i+L_i,\qquad i=1,2
    \end{align}
    Furthermore, we set
    \begin{align}
    L_1+L_2=I(V_1;V_2)+\epsilon
    \label{ensure_typical_pair_1}
    \end{align}
    to ensure that for given pairs $(w_1,\tilde{w}_1)$ and
    $(w_2,\tilde{w}_2)$, we can find a jointly typical pair
    $(\bold{v}_1(w_1,\tilde{w}_1,l_1),\bold{v}_2(w_2,\tilde{w}_2,l_2))$
    for some $l_1,l_2$.

    \item For each $(w_1,w_2)$, the transmitter randomly picks
    $(\tilde{w}_1,\tilde{w}_2)$ and finds a pair  $(\bold{v}_1(w_1,\tilde{w}_1,l_1),\break\bold{v}_2(w_2,\tilde{w}_2,l_2))$
    that is jointly typical. Such a pair exists with high
    probability due to (\ref{ensure_typical_pair_1}). Then, given
    this pair of $(\bold{v}_1,\bold{v}_2)$, the transmitter generates
    its channel inputs through $\prod_{i=1}^n p(x_i
    |v_{1,i},v_{2,i})$.

    \item User $j$ generates $2^{nR_{0,j}}$ length-$n$ sequences $\bold{u}_j$ through
    $p(\bold{u}_j)=\prod_{i=1}^{n}p(u_{j,i})$
    and labels them as $\bold{u}_j(s_{j,i})$ where $s_{j,i}\in
    \{1,\ldots,2^{nR_{0,j}}\}$ where $j=1,2$.

    \item For each $\bold{u}_j(s_{j,i})$, user $j$ generates
    $2^{n\hat{R}_j}$ length-$n$ sequences $\hat{\bold{y}}_j$ through
    $p(\hat{\bold{y}}_j|\bold{u}_j)=\prod_{i=1}^n
    p(\hat{y}_{j,i}|u_{j,i})$ and indexes them as $\hat{\bold{y}}_j
    (z_{j,i}|s_{j,i})$ where $z_{j,i}\in\{1,\ldots,2^{n\hat{R}_j}\},$ $j=1,2$.

    \item For each $\bold{u}_j(s_{j,i})$, user $j$ generates
    $2^{nR_{0,j}^{\prime}}$ length-$n$ sequences $\bold{x}_j$ through
    $p(\bold{x}_j|\bold{u}_j)=\prod_{i=1}^n p(x_{j,i}|u_{j,i})$ and
    indexes them as $\bold{x}_j(t_{j,i}|s_{j,i})$ where
    $t_{j,i}\in\{1,\ldots,2^{nR_{0,j}^{\prime}}\},$ $j=1,2$.

\end{enumerate}

\noindent\underline{\textbf{Partitioning:}}

\begin{itemize}
    \item Partition $2^{n\hat{R}_j}$ into cells $S_{s_{j,i}}$ where
$s_{j,i}\in\{1,\ldots,2^{nR_{0,j}}\},$ $j=1,2$.
\end{itemize}

\noindent\underline{\textbf{Encoding:}}

\vspace{0.2cm} The transmitter sends $\bold{x}$ corresponding to
the pair $(w_1,w_2)$. User $j$ sends $\bold{x}_j(t_{j,i}|s_{j,i})$
if the estimate of $\bold{y}_{j}(i-1)$, i.e., $\hat{z}_{j,i-1}$,
falls into $S_{s_{j,i}}$ and $t_{j,i}$ is chosen randomly from
$\{1,\ldots,2^{nR_{0,j}^{\prime}}\}$. The use of many
$\bold{x}_j(t_{j,i}|s_{j,i})$ for actual help signal
$\bold{u}_j(s_{j,i})$ aims to confuse the other user and to
decrease its decoding capability.

\noindent\underline{\textbf{Decoding:}}

\vspace{0.2cm}

We only consider decoding at user 1. Final expressions regarding
user 2 will follow due to symmetry.

\begin{enumerate}
    \item User 1 seeks a unique jointly typical pair of
    $\left(\bold{y}_1(i),\bold{u}_2(s_{2,i})\right)$
    which can be found with vanishingly small error probability if
    \begin{equation}
    R_{0,2}\leq I(U_2;Y_1|X_1)
    \end{equation}
    \item User 1 decides on $\bold{\hat{y}}_1({z_{1,i}|s_{1,i}})$
    by looking for a jointly typical pair $(\bold{\hat{y}}_1({z_{1,i}|s_{1,i}}),\bold{y}_1(i),\break\bold{u}_2(s_{2,i}),\bold{x}_1(t_{1,i}|s_{1,i}))$
    which can be ensured to exist if
    \begin{align}
    \hat{R}_1 \geq I(\hat{Y}_1;Y_1|U_1,U_2,X_1)
    \end{align}

    \item User 1 employs list decoding to decode
    $\bold{\hat{y}}_2(z_{2,{i-1}}|s_{2,{i-1}})$. It first calculates its
    ambiguity set as
    \begin{equation}
\mathcal{L}\left(\hat{\bold{y}}_2(z_{2,{i-1}}|\hat{s}_{2,{i-1}})\right)=\left\{\hat{\bold{y}}_2(z_{2,{i-1}}|\hat{s}_{2,{i-1}}):\left(\hat{\bold{y}}_2(z_{2,{i-1}}|\hat{s}_{2,{i-1}}),\bold{y}_{1}(i-1)\right)\textrm{ is jointly typical}  \right\}
    \end{equation}
    and then takes its intersection with $S_{\hat{s}_{2,i}}$ which results
    in a unique and correct intersection point if
    \begin{equation}
    \label{Compression_constraint_2_two}
    \hat{R}_2\leq
    I(\hat{Y}_2;Y_1|U_2,X_1)+R_{0,2}\leq I(\hat{Y}_2,U_2;Y_1|X_1)
    \end{equation}
    \item User 1 decides that $\bold{v}_1(w_{1,i-1},\tilde{w}_{1,i-1},l_{1,i-1})$ is received if
    there exists a unique jointly typical pair
    $\left(\bold{v}_1(w_{1,i-1},\tilde{w}_{1,i-1},l_{1,i-1}),\bold{y}_{1}(i-1),\hat{\bold{y}}_{2}(\hat{z}_{2,i-1}|\hat{s}_{2,i-1})\right)$
    which can be found with vanishingly small error probability
    if
    \begin{equation}
    \label{rate_2_new}
    R(V_1)\leq I(V_1;Y_1,\hat{Y}_2|X_1,U_2)
    \end{equation}
\end{enumerate}

\noindent\underline{\textbf{Equivocation computation:}}

\vspace{0.2cm}

Similar to the previous proofs, we treat each case separately. Due
to symmetry, we only consider user 1. If the rate of user 1 is
such that
\begin{align}
R_1\geq
I(V_1;Y_1,\hat{Y}_2|X_1,U_2)-I(V_1;Y_2,\hat{Y}_1|X_2,V_2,U_1)-I(V_1;V_2)
\end{align}
then we select the total number of codewords as
\begin{align}
R(V_1)=I(V_1;Y_1,\hat{Y}_2|X_1,U_2)
\label{total_number_codewords_two_sided_1}
\end{align}
which implies that
\begin{align}
\tilde{R}_1+L_1 \leq I(V_1;Y_2,\hat{Y}_1|X_2,V_2,U_1)+I(V_1;V_2)
\end{align}
The equivocation rate can be bounded as follows,
\begin{align}
H(W_1|Y_2^n,X_2^n)&\geq H(W_1|Y_2^n,X_2^n,\hat{Y}_1^n,V_2^n,U_1^n) \\
&=
H(W_1,Y_2^n,\hat{Y}_1^n,V_2^n|X_2^n,U_1^n)-H(Y_2^n,\hat{Y}_1^n,V_2^n|X_2^n,U_1^n)\\
&=
H(W_1,V_1^n,Y_2^n,\hat{Y}_1^n,V_2^n|X_2^n,U_1^n)-H(V_1^n|W_1,Y_2^n,\hat{Y}_1^n,V_2^n,X_2^n,U_1^n)\nonumber\\
&\quad -H(Y_2^n,\hat{Y}_1^n,V_2^n|X_2^n,U_1^n)\\
&=
H(V_1^n|X_2^n,U_1^n)+H(W_1,Y_2^n,\hat{Y}_1^n,V_2^n|X_2^n,U_1^n,V_1^n)\nonumber\\
&\quad -H(V_1^n|W_1,Y_2^n,\hat{Y}_1^n,V_2^n,X_2^n,U_1^n)-H(Y_2^n,\hat{Y}_1^n,V_2^n|X_2^n,U_1^n)\\
&\geq
H(V_1^n|X_2^n,U_1^n)-I(V_1^n;Y_2^n,\hat{Y}_1^n,V_2^n|X_2^n,U_1^n)\nonumber\\
&\quad -H(V_1^n|W_1,Y_2^n,\hat{Y}_1^n,V_2^n,X_2^n,U_1^n)
\label{equi_two_sided_case1_step1}
\end{align}
We treat each term in (\ref{equi_two_sided_case1_step1})
separately. The first term is
\begin{align}
H(V_1^n|X_2^n,U_1^n)=H(V_1^n)=nR(V_1)=nI(V_1;Y_1,\hat{Y}_2|X_1,U_2)
\label{equi_two_sided_case1_step2}
\end{align}
where the first equality is due to the independence of $V_1^n$ and
$(X_2^n,U_1^n)$, the second equality follows from the fact that
$V_1^n$ can take $2^{nR(V_1)}$ values with equal probability and
the last equality is due to our choice in
(\ref{total_number_codewords_two_sided_1}). The second term of
(\ref{equi_two_sided_case1_step1}) can be bounded as
\begin{align}
I(V_1^n;Y_2^n,\hat{Y}_1^n,V_2^n|X_2^n,U_1^n)\leq
nI(V_1;Y_2,\hat{Y}_1,V_2|X_2,U_1)+n\epsilon_n
\label{equi_two_sided_case1_step3}
\end{align}
following Lemma~3 of \cite{Ruoheng}. To bound the last term of
(\ref{equi_two_sided_case1_step1}), we consider the case that user
2 is trying to decode $V_1^n$ given the side information
$W_1=w_1$. Since $V_1^n$ can take
$2^{n(I(V_1;Y_2,\hat{Y}_1|X_2,V_2,U_1)+I(V_1;V_2))}$ values at
most, user 2 can decode $V_1^n$ with vanishingly small error
probability as long as this side information is available. Hence, the
use of Fano's lemma yields
\begin{align}
H(V_1^n|W_1,Y_2^n,\hat{Y}_1^n,V_2^n,X_2^n,U_1^n)\leq \epsilon_n
\label{equi_two_sided_case1_step4}
\end{align}
Plugging (\ref{equi_two_sided_case1_step2}),
(\ref{equi_two_sided_case1_step3}),
(\ref{equi_two_sided_case1_step4}) into
(\ref{equi_two_sided_case1_step1}), we get
\begin{align}
H(W_1|Y_2^n,X_2^n)&\geq
nI(V_1;Y_1,\hat{Y}_2|X_1,U_2)-nI(V_1;Y_2,\hat{Y}_1,V_2|X_2,U_1)-n\epsilon_n\\
&=nI(V_1;Y_1,\hat{Y}_2|X_1,U_2)-nI(V_1;Y_2,\hat{Y}_1|X_2,V_2,U_1)-nI(V_1;V_2)-n\epsilon_n
\label{independence_x2u1_v1v2}
\end{align}
where (\ref{independence_x2u1_v1v2}) follows from the independence
of $(X_2,U_1)$ and $(V_1,V_2)$, i.e., $I(V_1;V_2|X_2,U_1)\break =I(V_1;V_2)$.

For the other case, i.e., if the rate of user 1 is such that
\begin{align}
R_1\leq
I(V_1;Y_1,\hat{Y}_2|X_1,U_2)-I(V_1;Y_2,\hat{Y}_1|X_2,V_2,U_1)-I(V_1;V_2)
\end{align}
we select the total number of codewords as
\begin{align}
R(V_1)=R_1+I(V_1;Y_2,\hat{Y}_1|X_2,V_2,U_1)+I(V_1;V_2)
\end{align}
and following the same lines of computation, we can show that
\begin{align}
H(W_1|Y_2^n,X_2^n)\geq nR_1-n\epsilon_n
\end{align}
completing the proof.

\bibliography{IEEEabrv,references2}
\bibliographystyle{unsrt}

\end{document}